\documentclass[twocolumn,twocolappendix]{aastex631}

\usepackage{amsmath}
\shorttitle{CO emission surfaces in protoplanetary disks}
\shortauthors{Law et al.}
\graphicspath{{./}{figures/}}

\begin{document}

\title{CO Line Emission Surfaces and Vertical Structure in Mid-Inclination Protoplanetary Disks}

\author[0000-0003-1413-1776]{Charles J. Law}
\affiliation{Center for Astrophysics \textbar\, Harvard \& Smithsonian, 60 Garden St., Cambridge, MA 02138, USA}

\author[0000-0002-1050-278X]{Sage Crystian}
\affiliation{Center for Astrophysics \textbar\, Harvard \& Smithsonian, 60 Garden St., Cambridge, MA 02138, USA}

\author[0000-0003-1534-5186]{Richard Teague}
\affiliation{Center for Astrophysics \textbar\, Harvard \& Smithsonian, 60 Garden St., Cambridge, MA 02138, USA}

\author[0000-0001-8798-1347]{Karin I. \"Oberg}
\affiliation{Center for Astrophysics \textbar\, Harvard \& Smithsonian, 60 Garden St., Cambridge, MA 02138, USA}

\author[0000-0002-1779-8181]{Evan A. Rich}
\affiliation{Department of Astronomy, University of Michigan, 323 West Hall, 1085 South University Avenue, Ann Arbor, MI 48109, USA}

\author[0000-0003-2253-2270]{Sean M. Andrews}
\affiliation{Center for Astrophysics \textbar\, Harvard \& Smithsonian, 60 Garden St., Cambridge, MA 02138, USA}

\author[0000-0001-7258-770X]{Jaehan Bae}
\affiliation{Department of Astronomy, University of Florida, Gainesville, FL 32611, USA}

\author[0000-0003-2657-1314]{Kevin Flaherty} \affiliation{Department of Astronomy and Department of Physics, Williams College, Williamstown, MA 01267, USA}

\author[0000-0003-4784-3040]{Viviana V. Guzm\'{a}n}
\affiliation{Instituto de Astrof\'isica, Pontificia Universidad Cat\'olica de Chile, Av. Vicu\~na Mackenna 4860, 7820436 Macul, Santiago, Chile}

\author[0000-0001-6947-6072]{Jane Huang}
\altaffiliation{NASA Hubble Fellowship Program Sagan Fellow}
\affiliation{Department of Astronomy, University of Michigan, 323 West Hall, 1085 South University Avenue, Ann Arbor, MI 48109, USA}

\author[0000-0003-1008-1142]{John~D.~Ilee}
\affiliation{School of Physics and Astronomy, University of Leeds, Leeds, UK, LS2 9JT}

\author[0000-0002-3138-8250]{Joel H. Kastner}
\affiliation{Chester F. Carlson Center for Imaging Science, School of Physics \& Astronomy, and Laboratory for Multiwavelength Astrophysics, Rochester Institute of Technology, Rochester, NY 14623, USA}

\author[0000-0002-8932-1219]{Ryan A. Loomis}
\affiliation{National Radio Astronomy Observatory, 520 Edgemont Rd., Charlottesville, VA 22903, USA}

\author[0000-0002-7607-719X]{Feng Long}
\affiliation{Center for Astrophysics \textbar\, Harvard \& Smithsonian, 60 Garden St., Cambridge, MA 02138, USA}

\author[0000-0002-1199-9564]{Laura M. P\'erez}
\affiliation{Departamento de Astronom\'ia, Universidad de Chile, Camino El Observatorio 1515, Las Condes, Santiago, Chile}
\affiliation{N\'ucleo Milenio de Formaci\'on Planetaria (NPF), Chile}

\author[0000-0003-2953-755X]{Sebasti\'{a}n P\'{e}rez}
\affiliation{Center for Interdisciplinary Research in Astrophysics and Space Exploration (CIRAS), Universidad de Santiago de Chile}
\affiliation{Departamento de F\'{i}sica, Universidad de Santiago de Chile. Avenida Ecuador 3493, Estaci\'{o}n Central, Santiago, Chile}

\author[0000-0001-8642-1786]{Chunhua Qi}
\affiliation{Center for Astrophysics \textbar\, Harvard \& Smithsonian, 60 Garden St., Cambridge, MA 02138, USA}

\author[0000-0003-4853-5736]{Giovanni P. Rosotti}
\affiliation{Leiden Observatory, Leiden University, P.O. Box 9513, NL-2300 RA Leiden, the Netherlands}

\author[0000-0003-3573-8163]{Dary Ru\'{i}z-Rodr\'{i}guez}
\affiliation{National Radio Astronomy Observatory, 520 Edgemont Rd., Charlottesville, VA 22903, USA}

\author[0000-0002-6034-2892]{Takashi Tsukagoshi} \affiliation{National Astronomical Observatory of Japan, 2-21-1 Osawa, Mitaka, Tokyo 181-8588, Japan}

\author[0000-0003-1526-7587]{David J. Wilner}
\affiliation{Center for Astrophysics \textbar\, Harvard \& Smithsonian, 60 Garden St., Cambridge, MA 02138, USA}



\begin{abstract}
High-spatial-resolution CO~observations of mid-inclination~(${\approx}30$-75\degr)~protoplanetary disks offer an opportunity to study the vertical distribution of CO emission and temperature. The asymmetry of line emission relative to the disk major axis allows for a direct mapping of the emission height above the midplane, and for optically-thick, spatially-resolved emission in LTE, the intensity is a measure of the local gas temperature. Our analysis of ALMA archival data yields CO~emission surfaces, dynamically-constrained stellar host masses, and disk atmosphere gas temperatures for the disks around: HD~142666, MY~Lup, V4046~Sgr, HD~100546, GW~Lup, WaOph~6, DoAr~25, Sz~91, CI~Tau, and DM~Tau. These sources span a wide range in stellar masses~(0.50-2.10~M$_{\odot}$), ages~(${\sim}$0.3-23~Myr), and CO~gas radial emission extents~(${\approx}$200-1000~au). This sample nearly triples the number of disks with mapped emission surfaces and confirms the wide diversity in line emitting heights ($z/r\approx0.1$~to~${\gtrsim}0.5$) hinted at in previous studies. We compute radial and vertical CO~gas temperature distributions for each disk. A few disks show local temperature dips or enhancements, some of which correspond to dust substructures or the proposed locations of embedded planets. Several emission surfaces also show vertical substructures, which all align with rings and gaps in the millimeter dust. Combining our sample with literature sources, we find that CO~line emitting heights weakly decline with stellar mass and gas temperature, which, despite large scatter, is consistent with simple scaling relations. We also observe a correlation between CO~emission height and disk size, which is due to the flared structure of disks. Overall, CO~emission surfaces trace~${\approx}2$-$5\times$~gas pressure scale heights~(H$_{\rm{g}}$)~and could potentially be calibrated as empirical tracers of~H$_{\rm{g}}$.
\end{abstract}

\textbf{}\\\\
\section{Introduction} \label{sec:intro}

Protoplanetary disks exhibit flared emitting surfaces set by hydrostatic equilibrium, as first recognized in the spectral energy distributions of their host stellar systems \citep{Kenyon87}. Disks are also highly stratified in their physical and chemical properties \citep{Williams11} with vertical distributions of molecular material that are greatly influenced by gradients in physical conditions such as gas temperature, density, or radiation \citep[e.g.,][]{Walsh10, Fogel11}, the efficiency of turbulent vertical mixing \citep[e.g.,][]{Ilgner04, Semenov11,Flaherty20}, or the presence of meridional flows driven by embedded planets \citep[e.g.,][]{Morbidelli14, Teague19Natur, Yu21}.

A detailed understanding of this complex vertical structure is required to interpret kinematic signals in CO emission \citep{Perez15_gas_planets, Perez18, Pinte19Nat, DiskDyn20, Perez20, Teague21} and the effects of embedded protoplanets on the density distribution, temperature, and pressure of gas in disks \citep{Teague18_AS209, Calcino21, Izquierdo21}. Accurate dynamical mass estimates derived from line emission rotation maps also require well-constrained line emitting heights \citep{Casassus19, Veronesi21}. This is especially critical as most line emission does not originate from the midplane but from layers higher up in the disk \citep{Dartois03, Pietu07}. As a result, line emission surfaces also trace the vertical temperature structure of disks \citep{Dartois03, Rosenfeld13, pinte18, Teague20_goham, Law21, Flores21}, provide important inputs to disk thermochemical models \citep{Zhang21, Calahan21, Schwarz21}, and serve as useful diagnostics to disentangle observational signatures of planet-disk interactions versus depletions in gas surface density \citep{Dong19, Rab20, Bae21, Alarcon21}. Emission surfaces are also relevant for the chemistry of planet formation, as they are required to assess how well connected molecular gas abundances derived from line observations are to their abundances in planet-forming disk midplanes.


There are several approaches to obtaining information about the vertical distribution of gas in disks. Vertical structures have been observed in highly-inclined or edge-on disks, which allow a direct mapping of emission distributions \citep[e.g.,][]{Guilloteau16, Dutrey17, Teague20_goham, Podio20, RR21, Flores21, Villenave22}. However, with sufficient angular resolution and surface brightness sensitivity it is possible to extract disk vertical structures from mid-inclination (${\approx}$30--75\degr) disks by exploiting spatially-resolved emission from elevated regions above and below the midplane \citep[e.g.,][]{Gregorio13,Rosenfeld13,Isella18}. In these cases, the emission heights of bright molecular lines can be directly determined \citep{pinte18, Rich21, Paneque21, Law21}. This approach expands the sample of disks whose vertical structure can be mapped and allows us to relate vertical gas structure to that of the radial continuum, which is often inaccessible in edge-on disks due to high optical depths.




With a known temperature structure, it is also possible to estimate indirect line emission heights based on inferred brightness temperatures for disks with low inclinations \citep[e.g.,][]{Teague20, Oberg20_Rosetta}, or for molecules with weaker emission where direct mapping is not feasible \citep[e.g.,][]{Ilee21}. Without such a temperature structure, relative stratification patterns between different molecular emission lines can be discerned by modeling multiple line fluxes \citep[e.g.,][]{Bruderer12,Fedele16}.

\begin{figure*}[p!]
\centering
\includegraphics[width=\linewidth]{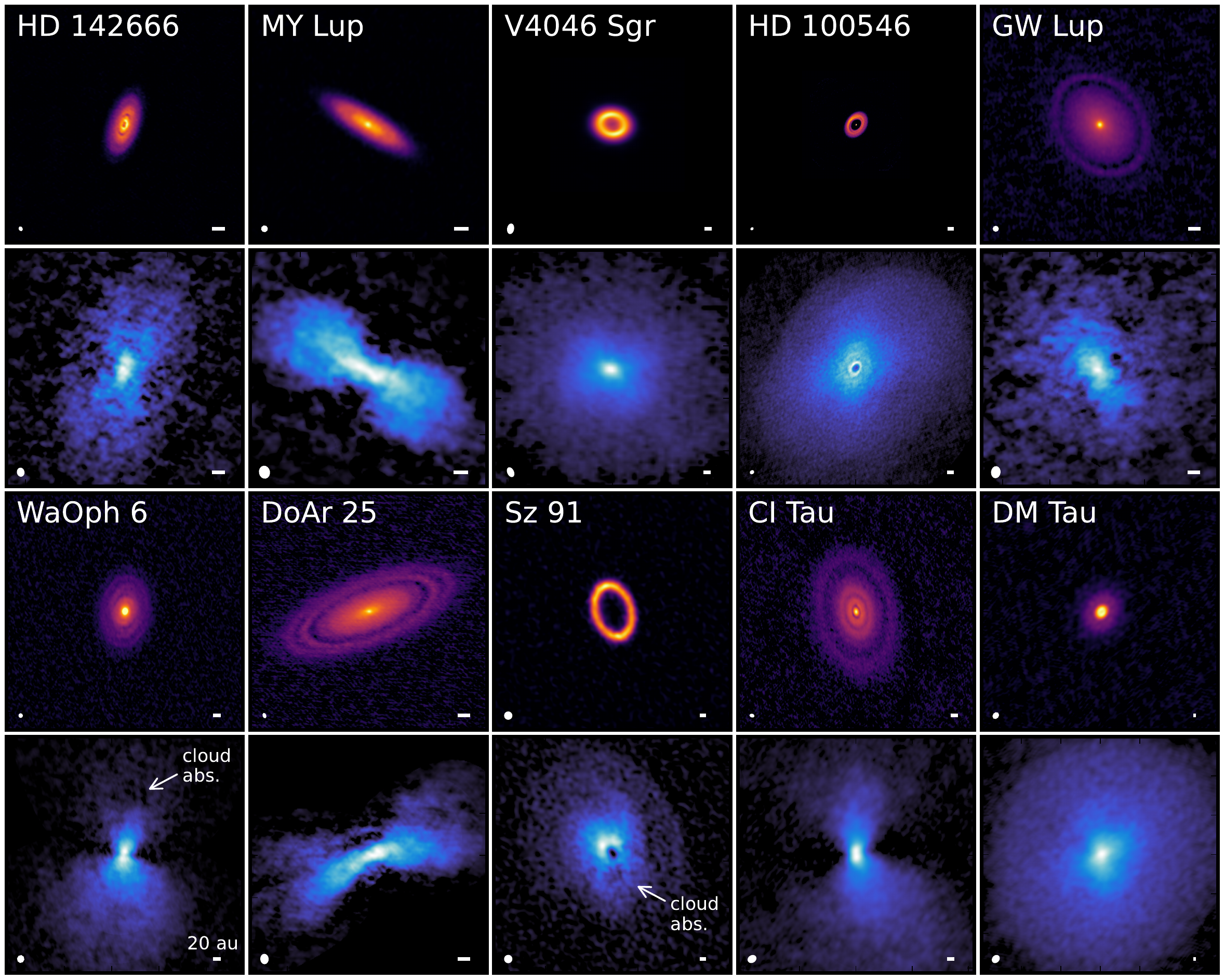}
\caption{Millimeter continuum images (first and third rows) and CO zeroth moment maps (second and fourth rows) for all disks. Line emission is from CO J=2--1 and the continuum is at 1.3~mm, except for V4046~Sgr and Sz~91, which show the 870~$\mu$m continuum; and for V4046~Sgr, Sz~91, and CI~Tau, which show CO J=3--2 line emission. Panels for each disk have the same field of view. Color stretches were individually optimized and applied to each panel to increase the visibility of outer disk structure. The asymmetries present in WaOph~6 and Sz~91 are due to cloud contamination and are labeled in the maps of each disk. The dark lane seen in DoAr~25 traces the disk midplane and is visible due to the relatively high inclination of this source. The synthesized beam and a scale bar indicating 20~au is shown in the lower left and right corner, respectively, of each panel. Details about each of the observations are found in Section \ref{sec:archival_data} and Table \ref{tab:disk_char}.}
\label{fig:figure1}
\end{figure*}

As part of the Molecules with ALMA at Planet-forming Scales (MAPS) \citep{Oberg21_MAPSI} ALMA Large Program, \citet{Law21} directly mapped the emission surfaces of several CO isotopologues in the disks around IM~Lup, GM~Aur, AS~209, HD~163296, and MWC~480. The authors found a wide range in CO line emitting heights and identified tentative trends suggesting that disks with lower host star masses and larger CO gas disks had more vertically extended emission surfaces. However, firm conclusions were precluded by the small sample size of five disks.

Here, we extract CO emission surfaces for ten disks with favorable orientations with respect to our line-of-sight that have been previously observed at sufficiently high spatial resolution and sensitivity. We describe the ALMA archival data from which we draw our disk sample and briefly detail our surface extraction methods in Section \ref{sec:observations_overview}. In Section \ref{sec:results}, we present the derived emission surfaces, compare them with previous millimeter and NIR observations, and calculate radial and vertical temperature profiles. We explore possible origins of the observed disk vertical structures and examine the relationship between line emission surfaces and gas pressure scale heights in Section \ref{sec:discussion}. We summarize our conclusions in Section \ref{sec:conlcusions}.

\section{Observations and analysis}
\label{sec:observations_overview}

\subsection{Archival Data}
\label{sec:archival_data}

We searched the ALMA archive for CO line observations of protoplanetary disks with inclinations of 30-75\degr and sufficiently high angular resolutions, line sensitivities, and velocity resolutions to derive emission surfaces.

We made use of the publicly available, science ready CO J=2--1 image cubes from the ALMA Large Program DSHARP\footnote{\url{https://bulk.cv.nrao.edu/almadata/lp/DSHARP/}} \citep{Andrews18}. We selected those disks with favorable inclinations for surface extractions and excluded those disks with prohibitively severe cloud contamination. After these considerations, we were left with the following sources: HD~142666, MY~Lup, GW~Lup, WaOph~6, and DoAr~25. We also excluded the disks observed as part of MAPS, as they already have well-constrained emission surfaces \citep{Law21}. In addition, we used ALMA observations of the disks around: V4046~Sgr \citep{RuizRodriguez19}, HD~100546 \citep{Perez20}, Sz~91 \citep{Tsukagoshi19}, CI~Tau \citep{Rosotti21}, and DM~Tau \citep{Flaherty20}. All data were obtained from the original authors and observational details may be found in the corresponding references. The data for V4046~Sgr, Sz~91, and CI~Tau are CO J=3--2, while DM~Tau and HD~100546 are CO J=2--1. Velocity resolutions spanned from 0.16-0.5~km~s$^{-1}$, while typical angular resolutions were ${\approx}$0\farcs07--0\farcs14, or 10-20~au, with the exception of DM~Tau (0\farcs36; 52~au). The large size of the DM~Tau CO gas disk and its highly flared nature \citep[e.g.,][]{Flaherty20} made surface extraction possible even with a coarser angular resolution.

Overall, the sources in our sample span a wide range in both stellar properties, such as masses (0.50-2.10~M$_{\odot}$), spectral types (M-B), bolometric luminosities (0.24-23.4~L$_{\odot}$), and ages (${\sim}$0.3-23~Myr), as well as disk physical characteristics, such as CO gas disk radial emission extents (${\approx}$200-1000~au), and includes both full and transition disks. Several of our sources exhibit mild-to-moderate cloud contamination, in which the ambient cloud significantly absorbs disk line emission with overlapping velocities. This is identified through visual inspection of channel maps and manifests as spatial brightness asymmetries in images of the CO line emission. Table \ref{tab:disk_char} shows a summary of source characteristics, including the ALMA Project Codes for the corresponding archival data.

Figure \ref{fig:figure1} shows an overview of the disk sample in millimeter continuum emission and CO velocity-integrated intensity, or ``zeroth moment," maps. All continuum images are taken from previously published ALMA observations. Specifically, we show 1.3~mm continuum images of HD~142666, MY~Lup, GW~Lup, WaOph~6, and DoAr~25 \citep[][]{Andrews18}; HD~100546 \citep{Perez20}; CI~Tau \citep{Clarke18}, and DM~Tau \citep{Flaherty20}. We show 870~$\mu$m continuum images of V4046~Sgr \citep{RuizRodriguez19} and Sz~91 \citep{Canovas16}. We generated the zeroth moment maps from the CO image cubes using \texttt{bettermoments} \citep{Teague18_bettermoments} with no sigma clipping and Keplerian masks based on the parameters in Table \ref{tab:disk_char}. See Appendix \ref{sec:appendix_disksize_profiles} for more details on the moment map generation process.

For the calculation of gas temperatures with the full Planck function, we also made use of the line+continuum image cubes. These were also obtained from the original authors, with the exception of the DSHARP sources, where we manually re-imaged the line emission cubes with the continuum following the same imaging procedures used to produce the original CO cubes \citep{Andrews18}. We also re-imaged archival data (PI: G. van der Plas, 2015.1.00192.S) of the HD~97048 disk to derive a line+continuum image cube (Appendix \ref{sec:appendix_HD97048}). This source is not formally part of our sample as it already has a directly-mapped CO line emission surface from \citet{Rich21} but lacks an estimate of its CO gas temperature structure. While the CO thermal structure of the HD~97048 disk is of interest in its own right, it is also required for establishing a homogeneous sample for source-to-source comparisons.

The line-only and line+continuum image cubes as well as all zeroth moment maps are publicly available on Zenodo doi: 10.5281/zenodo.6410045.

\begin{deluxetable*}{lccccccccccc}[ht!]
\tablecaption{Stellar and Disk Characteristics\label{tab:disk_char}}
\tablewidth{0pt}
\tablehead{
\colhead{Source} & \colhead{Spectral} & \colhead{Distance\tablenotemark{a}} & \colhead{incl.} & \colhead{PA} & \colhead{M$_*$\tablenotemark{b}} &
\colhead{L$_*$} & \colhead{Age\tablenotemark{c}} & \colhead{v$_{\rm{sys}}$\tablenotemark{b}} & \colhead{cloud} & \colhead{ALMA} & \colhead{Ref.}\vspace{-0.2cm}\\
\colhead{} & \colhead{Type} & \colhead{(pc)} & \colhead{($^{\circ}$)}  & \colhead{($^{\circ}$)} & \colhead{(M$_{\odot}$)} & \colhead{(L$_{\odot}$)} & \colhead{(Myr)} & \colhead{(km~s$^{-1}$)}  & \colhead{contam.} & \colhead{Project Code} & \colhead{}}
\startdata
HD~142666 & A8 & 145 & 62.2 & 162.1 & 1.73 & 9.1 & 13 & 4.37 & \ldots & 2016.1.00484.L & 1,2 \\
MY~Lup & K0 & 157 & 73.2 & 58.8 & 1.27 & 0.87 & 10 & 4.71 & mild & 2016.1.00484.L & 1,2 \\
V4046~Sgr\tablenotemark{d} & K5,K7 & 71 & 34.7 & 75.7 & 1.72 & 0.86 & 23 & 2.93 & \ldots & 2016.1.00315.S & 3-8 \\
HD~100546 & B9 & 108 & 41.7 & 146.0 & 2.10 & 23.4 & 5 & 5.65 & \ldots & 2016.1.00344.S & 9-13 \\
GW~Lup & M1.5 & 154 & 38.7 & 37.6 & 0.62 & 0.33 & 2 & 3.69 & \ldots & 2016.1.00484.L & 1,2 \\
WaOph~6 & K6 & 122 & 47.3 & 174.2 & 1.12 & 2.9 & 0.3 & 4.21 &  mild & 2016.1.00484.L & 1,2 \\
DoAr~25 & K5 & 138 & 67.4 & 110.6 & 1.06 & 0.95 & 2 & 3.38 & moderate & 2016.1.00484.L & 1,2 \\
Sz~91 & M0 & 158 & 49.7 & 18.1 & 0.55 & 0.26 & 3-7 & 3.42 & moderate & 2012.1.00761.S & 14-16 \\
CI~Tau & K5.5 & 160 & 49.2 & 11.3 & 1.02 & 1.26 & 2 & 5.70 & moderate & 2017.A.00014.S & 17-20 \\
DM~Tau & M1 & 143 & 36.0 & 154.8 & 0.50 & 0.24 & 1-5 & 6.04 & \ldots & 2016.1.00724.S & 4,21-22\\
HD~97048 & A0V & 184 & 41.0 & 3.0 & 2.70 & 44.2 & 4 & 4.55 & moderate & 2015.1.00192.S & 23-25\\
\enddata
\tablenotetext{a}{All distances are from \textit{Gaia} DR3 \citep{Gaia21, Bailer_Jones21}.}
\tablenotetext{b}{Dynamical masses and systemic velocities are derived in this work (see Section \ref{sec:appendix_vkep}).}
\tablenotetext{c}{Stellar ages are likely uncertain by at least a factor of two.}
\tablenotetext{d}{V4046~Sgr hosts a protoplanetary disk orbiting a binary star system. The individual stellar spectral types are listed, along with the total stellar mass and luminosity.}
\tablecomments{References are: 1. \citet{Andrews18}; 2. \citet{Huang18}; 3. \citet{Quast00}; 4. \citet{Flaherty20}; 5. \citet{Rosenfeld12}; 6. \citet{Mamajek14}; 7. \citet{Torres06}; 8. \citet{Binks14}; 9. \citet{Pineda14}; 10. \citet{Pineda19}; 11. \citet{Vioque18}; 12. \citet{Fedele21}; 13. \citet{Casassus19}; 14. \citet{Romero12}; 15. \citet{Tsukagoshi19}; 16. \citet{Mauco20}; 17. \citet{Clarke18}; 18. \citet{Simon17}; 19. \citet{Donati20}; 20. \citet{Simon19}; 21. \citet{Guilloteau14}; 22. \citet{Ancker98}; 23. \citet{Walsh16}; 24. \citet{vanderPlas17}; 25. \citet{Asensio_Torres21}.}
\end{deluxetable*}

\vspace{-25pt}
\subsection{Methods}
\label{sec:methods}

\subsubsection{Surface Extraction}
\label{sec:methods_sub_surfextr}
\vspace{-25pt}

\begin{deluxetable*}{lcccccc}[ht!]
\tablecaption{Parameters for CO Emission Surface Fits\label{tab:emission_surf}}
\tablewidth{0pt}
\tablehead{
\colhead{Source} & \colhead{Line} &  \multicolumn5c{Exponentially-Tapered Power Law} \\ \cline{3-7}
\colhead{} &  \colhead{} & \colhead{r$_{\rm{fit,\,max}}$ [$^{\prime \prime}$]} & \colhead{$z_0$ [$^{\prime \prime}$]} & \colhead{$\phi$} &\colhead{$r_{\rm{taper}}$ [$^{\prime \prime}$]} &\colhead{$\psi$}}
\startdata
HD~142666 & J=2$-$1 & 0.80 & 0.09$^{+0.12}_{-0.03}$ & 0.50$^{+0.44}_{-0.27}$ & 1.13$^{+0.59}_{-0.48}$ & 2.37$^{+5.10}_{-1.79}$\\
MY~Lup & J=2$-$1 & 1.00 & 0.21$^{+0.14}_{-0.08}$ & 1.28$^{+0.40}_{-0.41}$ & 0.80$^{+0.08}_{-0.10}$ & 3.95$^{+2.10}_{-1.29}$\\
V4046~Sgr & J=3$-$2 & 2.25 & 0.28$^{+0.08}_{-0.03}$ & 0.59$^{+0.23}_{-0.16}$ & 1.99$^{+0.18}_{-0.34}$ & 2.59$^{+1.22}_{-0.92}$\\
HD~100546 & J=2$-$1 & 1.20 & 0.35$^{+0.21}_{-0.07}$ & 1.09$^{+0.29}_{-0.18}$ & 1.02$^{+0.08}_{-0.20}$ & 2.57$^{+0.95}_{-0.79}$\\
GW~Lup & J=2$-$1 & 1.20 & 0.22$^{+0.07}_{-0.02}$ & 0.76$^{+0.19}_{-0.14}$ & 1.22$^{+0.21}_{-0.11}$ & 5.91$^{+2.48}_{-3.71}$\\
WaOph~6 & J=2$-$1 & 1.40 & 0.37$^{+0.24}_{-0.10}$ & 1.77$^{+0.36}_{-0.30}$ & 1.13$^{+0.15}_{-0.25}$ & 2.52$^{+1.27}_{-0.78}$\\
DoAr~25 & J=2$-$1 & 1.95 & 0.31$^{+0.02}_{-0.01}$ & 1.54$^{+0.13}_{-0.12}$ & 1.61$^{+0.03}_{-0.04}$ & 5.85$^{+1.05}_{-0.86}$\\
Sz~91\tablenotemark{a} & J=3$-$2 & 1.60 & 0.91$^{+0.07}_{-0.12}$ & 2.59$^{+0.12}_{-0.15}$ & 0.86$^{+0.06}_{-0.03}$ & 1.99$^{+0.18}_{-0.16}$\\
CI~Tau & J=3$-$2 & 1.40 & 0.32$^{+0.07}_{-0.03}$ & 1.48$^{+0.16}_{-0.14}$ & 2.07$^{+0.58}_{-0.31}$ & 2.61$^{+0.97}_{-1.06}$\\
DM~Tau & J=2$-$1 & 3.00 & 0.82$^{+0.06}_{-0.05}$ & 1.85$^{+0.08}_{-0.07}$ & 1.79$^{+0.11}_{-0.12}$ & 1.67$^{+0.10}_{-0.10}$\\
\textit{HD~97048}\tablenotemark{b} & \textit{J=2$-$1} & \textit{2.65} & \textit{0.31$^{+0.02}_{-0.01}$} & \textit{1.16$^{+0.09}_{-0.07}$} & \textit{2.74$^{+0.09}_{-0.13}$} & \textit{2.81$^{+0.54}_{-0.50}$}\\
\enddata
\tablenotetext{a}{Fit only considering the inner 1\farcs60 to avoid elevated, diffuse material at larger radii, which is not well-fit by an exponentially-tapered power law.}
\tablenotetext{b}{CO line emission surface rederived and fit with an exponentially-tapered power law for consistency. See Appendix \ref{sec:appendix_HD97048} and \citet{Rich21}.}
\end{deluxetable*}

\begin{figure*}[th!]
\centering
\includegraphics[width=\linewidth]{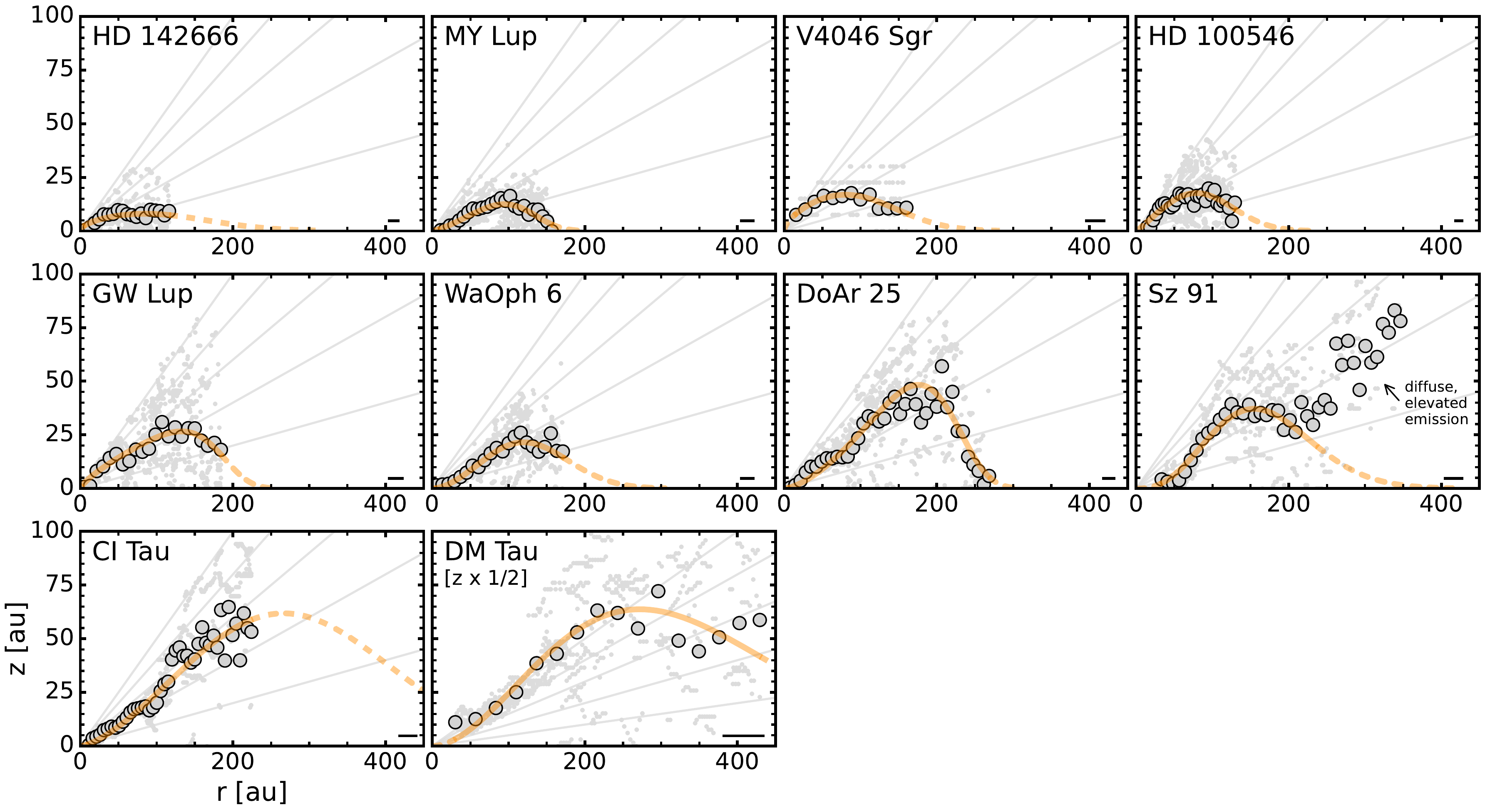}
\caption{CO emission surfaces for all disks. Large gray points show radially-binned surfaces and small, light gray points represent individual measurements. The orange lines show the exponentially-tapered power law fits from Table \ref{tab:emission_surf}. The solid lines show the radial range used in the fitting, while the dashed lines are extrapolations. Diffuse, elevated emission present at large radii in the Sz~91 disk is labeled and excluded in the fits. Lines of constant $z/r$ from 0.1 to 0.5 are shown in gray. All panels show a consistent radial and vertical range, except for DM~Tau where the vertical extent has been scaled by $\times$1/2. The FWHM of the major axis of the synthesized beam is shown in the bottom right corner of each panel. The emission surfaces shown in this figure are available as Data behind the Figure.}
\label{fig:figure_gallery_r_v_z}
\end{figure*}

We used the line emission image cubes to extract vertical emission surfaces for each disk, closely following the methods of \citet{Law21}. In short, we leveraged the spatially-resolved emission asymmetry visible in the channel maps (see Figure \ref{fig:isovel_contours}, Appendix \ref{sec:appendix_isovel}) to constrain the vertical emission height. To do so, we used the \texttt{disksurf} \citep{disksurf_Teague} python code, which implements this method as well as several filtering steps to extract more accurate emission surfaces.

For each image cube, we restricted the position-position-velocity regions from which we extracted surfaces to those contained in disk-specific Keplerian masks based on CO emission morphology and source characteristics. We then manually excluded those channels where the front and back disk sides could not be disentangled as well as those channels with severe cloud contamination. After the initial extraction, we filtered pixels based on priors of disk physical structure. We removed those pixels with extremely high $z$/$r$ values (upper boundaries ranging from 0.45 to 1.0 depending on the disk) and large negative $z$ values, as the emission must arise from at least the midplane. We allowed points with small negative values, i.e., $z$/$r > -0.1$, to remain to avoid positively biasing our averages to non-zero $z$ values. To minimize contamination from background thermal noise, which can confuse the identification of emission peaks, we also filtered points based on surface brightness thresholds, which varied from 1$\times$rms (HD~142666) to 8$\times$rms (DM~Tau). The wide range in thresholds was a result of our heterogeneous sample with differing line sensitivities, which was driven in part by varied beam sizes. For instance, the beam size of the DM~Tau observations is approximately five times greater than that of the HD~142666 image cubes. This is comparable to the source size ratio between the two disks, i.e., the DM~Tau disk is nearly five times larger than that of HD~142666. In general, we prioritized the extraction of the maximum number of reliable emission surface pixels and visually confirmed the quality of each extraction before and after the filtering process. For further details about this procedure, see \citet{Law21}.

Emission surfaces were extracted on a per-pixel basis. We first must assume an inclination and position angle of each disk (Table \ref{tab:disk_char}). Then, for each pixel associated with the emitting surface, we obtained a deprojected radius $r$, emission height $z$, surface brightness $I_{\nu}$, and channel velocity $v$. To further reduce scatter in these surfaces, we used two different binning methods: (1) we radially binned the surfaces using bins equal to one-half of the FWHM of the beam major axis; (2) and calculated a moving average with a minimum window size of 1/2$\times$ the beam major axis FWHM. The binned surfaces resulted in a uniform radial sampling, while the moving averages retained a finer radial sampling, which is essential for identifying subtle vertical perturbations in the emission surfaces that may be, e.g., associated with features in the dust continuum or putative planet locations. These are the same binning methods employed in \citet{Law21}, but with twice as large a radial bin and window size, due to the generally less sensitive data used here relative to that of the MAPS sample \citep{Oberg21_MAPSI}.

All three types of line emission surfaces -- individual measurements, radially-binned, and moving averages -- are made publicly available. Throughout this work, we sometimes radially bin these data products further for visual clarity, but all quantitative analysis is done using the original binning of each type of emission surface.

\subsubsection{Analytical Fitting}
\label{sec:methods_sub_fitting}

To more readily compare with other observations and to facilitate their incorporation into models, we fitted exponentially-tapered power laws to all CO emission surfaces. This fit describes both the flared surfaces in the inner disk and the plateau/turnover region in the outer disk. We adopt the same functional form as in \citet{Law21}:

\begin{equation} \label{eqn:exp_taper}
z(r) = z_0 \times \left( \frac{r}{1^{\prime \prime}} \right)^{\phi} \times \exp \left(- \left[ \frac{r}{r_{\rm{taper}}} \right]^{\psi} \right)
\end{equation}

where $z_0$, $\phi$, and $\psi$ are non-negative. A value of $\phi > 1$ indicates that $z/r$ increases with radius, while $0 < \phi < 1$ tends toward a flat $z(r)$ profile. 

All fits were performed using the Monte Carlo Markov Chain (MCMC) sampler implemented in \texttt{emcee} \citep{Foreman_Mackey13} to estimate the posterior distributions of the following parameters: $z_0$, $\phi$, r$_{\rm{taper}}$, and $\psi$. The radial range of each fit is given by r$_{\rm{fit,\,max}}$ in Table \ref{tab:emission_surf}. We used 64 walkers which take 1000 steps to burn in and an additional 500 steps to sample the posterior distribution function. We chose an MCMC fitting approach rather than a simple $\chi^2$ minimization, as we found that it better handled the degeneracies between fitted parameters, especially, e.g., between $\psi$ and r$_{\rm{taper}}$. Table \ref{tab:emission_surf} shows all fitted parameters. Isovelocity contours generated using the surface fits from Table \ref{tab:emission_surf} are shown in Figure \ref{fig:isovel_contours} in Appendix \ref{sec:appendix_isovel}.

\section{Results} \label{sec:results}

\subsection{Overview of Emission Surfaces} \label{sec:overview_emission_surfaces}

Figure \ref{fig:figure_gallery_r_v_z} shows the CO emission surfaces derived in all disks in our sample. There is considerable disk-to-disk variation in line emitting heights and surface flaring, i.e., how quickly $z$ increases as a function of $r$. Peak emitting heights range from ${\approx}$10-150~au, while typical $z$/$r$ values span ${\approx}$0.1 to ${\gtrsim}$0.5. HD~142666 hosts the flattest disk, while the DM~Tau disk has by far the most elevated emission surface.

Many of the disks exhibit a quick, power-law-like rise in height with radius, which is then followed by a gradual flattening and eventual turnover of their emission surfaces at large radii as, presumably, gas surface densities decrease. However, we sometimes only see either the initial flattening, like in the HD~142666 disk, or the beginning of the turnover phase, such as for the WaOph~6 and DM~Tau disks. We suspect that the missing turnovers are simply due to low SNR in the outer regions of some disks. For sources (e.g., CI~Tau) where the turnover is not visible, the r$_{\rm{taper}}$ and $\psi$ parameters of the analytical fits in Table \ref{tab:emission_surf} are highly uncertain.

Notably, the Sz~91 emission surface does not follow this characteristic structure. While we see the flared and plateau phases out to 200~au, emission heights again begin to quickly rise beyond this and do not show any sign of flattening out to ${\approx}$350~au. The presence of diffuse emission at large radii in this disk was previously noted by \citet{Tsukagoshi19}, and the derived surface is quite similar to that of CO J=2--1 in the IM~Lup disk \citep{Law21}. When fitting this disk, we thus restrict our analytic fits to within ${\approx}$200~au.

\begin{figure*}[h!]
\centering
\includegraphics[width=\linewidth]{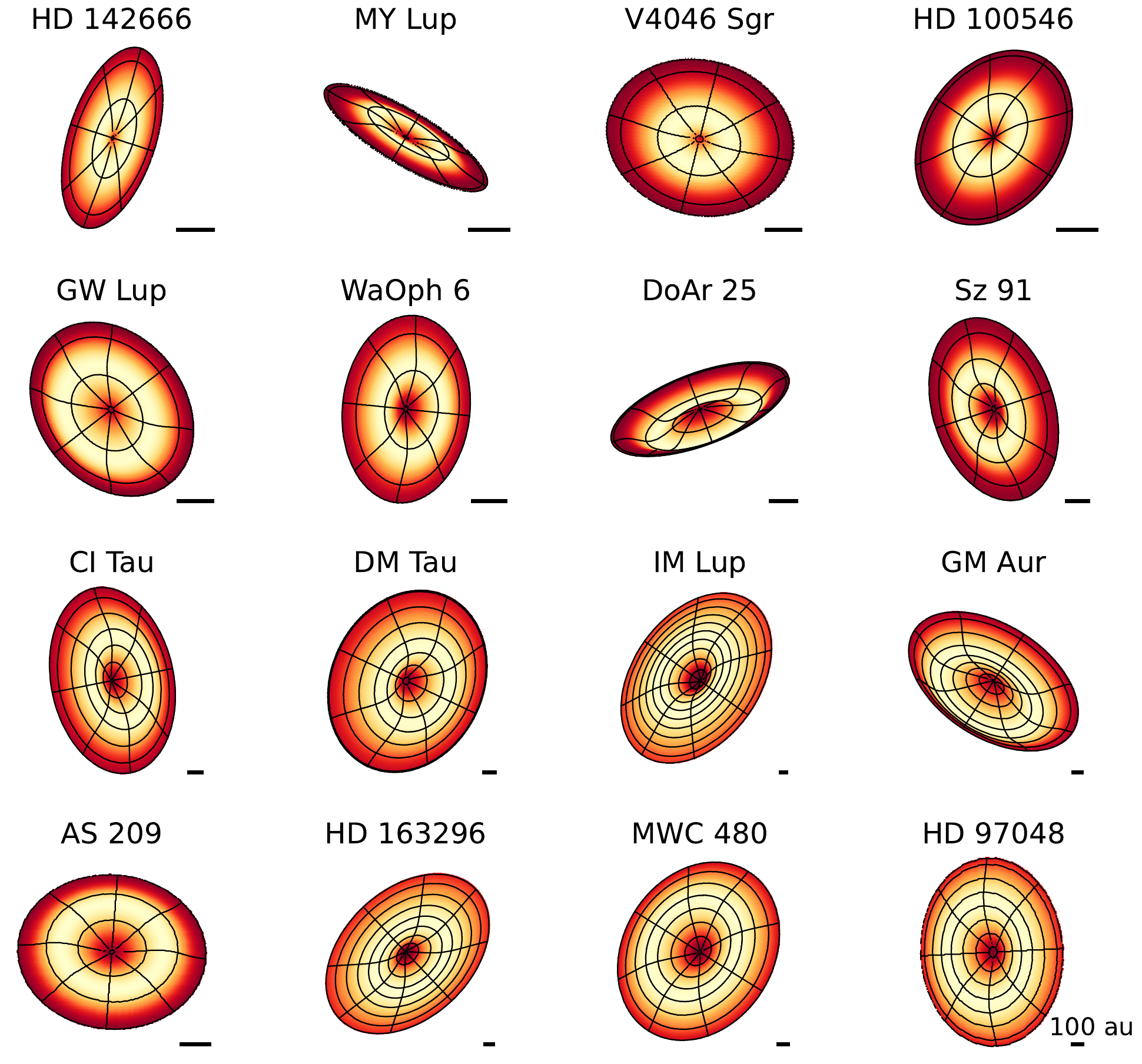}
\caption{Three-dimensional representations of CO emission surfaces for the disks derived in this work and for the disks around IM~Lup, GM~Aur, AS~209, HD~163296, and MWC~480 \citep{Law21} and HD~97048 \citep{Rich21}. Colormaps show the vertical height of each emission surface using exponentially-tapered power law profiles. For each disk, the colormap is normalized to the maximum height and each contour represents a radial distance of 100~au. Surfaces are radially extrapolated beyond the direct surface measurements in Figure \ref{fig:figure_gallery_r_v_z} to better illustrate their shapes; however, we caution that this sometimes results in a surface that is larger than the total CO gas disk extent. The elevated, diffuse emission at large radii in the Sz~91 and IM~Lup disks are not shown. A scale bar indicating 100~au is shown in the lower right corner.} 
\label{fig:3D_surfaces}
\end{figure*}

Overall, there is no single characteristic height that all disks share, but instead line emission heights vary by over an order of magnitude, while typical $z/r$ values span at least a factor of five. These results confirm that the diversity previously observed in line emission heights \citep{Law21} is commonplace. To better illustrate this and highlight the geometry of the emission surfaces, Figure \ref{fig:3D_surfaces} shows a 3D representation of the fitted surfaces in our disk sample and from literature sources with directly-mapped CO emission surfaces.

\subsection{Vertical Substructures and Comparison with Millimeter Continuum Features and Kinematic Planetary Signatures} \label{sec:vertical_substr_vs_mm_cont}

\begin{figure*}[]
\centering
\includegraphics[width=\linewidth]{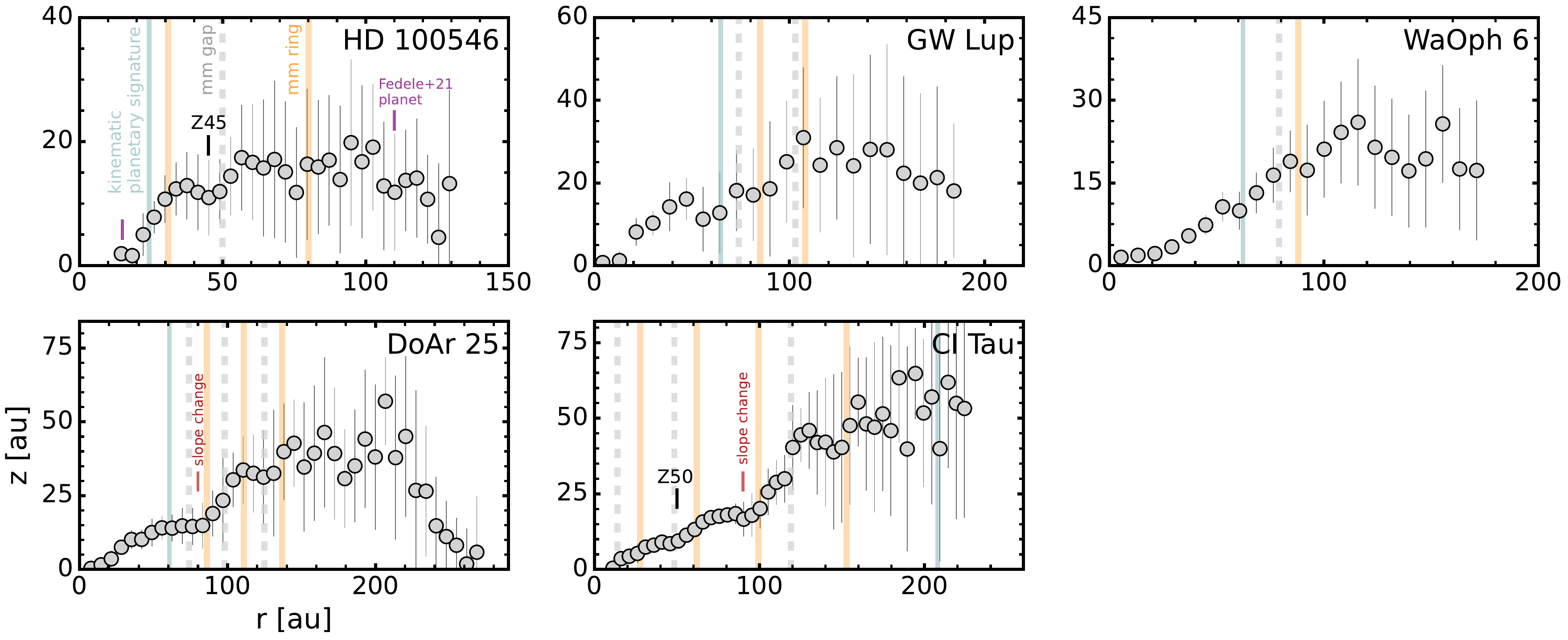}
\caption{CO emission surfaces for disks with vertical substructures or kinematic planetary signatures. Large gray points show radially-binned surfaces and vertical lines show the 1$\sigma$ uncertainties in $z$. Vertical substructures in CO emission surfaces are labeled by their approximate radial location in au following the nomenclature of \citet{Law21} and are marked in black, while slope changes are shown in red. The midpoint radial locations of millimeter dust rings and gaps are shown as solid orange and dashed gray lines, respectively, and are compiled from \citet{Huang18, Clarke18, Long18, Fedele21}. Radial locations of dust features in the HD~100546 disk are approximate, due to the azimuthally asymmetric dust emission in this source \citep{Pineda19, Perez20, Fedele21}. The mm~gap at 40~au in the HD~100546 disk marks the inner edge of a wide (${\sim}$40-150~au) continuum gap. KPSs are marked by blue lines and are from \citet{Casassus19, Perez20, Pinte20, Rosotti21}. The proposed radial locations of two Jupiter-mass planets (one at 15~au and another at 110~au) in the HD~100546 disk inferred from the smoothed-particle-hydrodynamic simulations of \citet{Fedele21} are shown in purple.}
\label{fig:substr_plot}
\end{figure*}

A few of the emission surfaces in our sample exhibit vertical substructures in the form of dips or prominent changes in emission slope. In Figure \ref{fig:substr_plot}, a dip at 45~au is evident in the line emitting heights of the HD~100546 disk, while slope changes are seen around 80~au and 90~au in the emission surfaces of the DoAr~25 and CI~Tau disks, respectively. A shallow dip is also seen at 50~au in the emission surface of the CI~Tau disk. 

Each of these vertical substructures radially aligns with dust features. In Figure \ref{fig:substr_plot}, we overlay the midpoint radial locations of millimeter rings and gaps in all disks. The radial locations of dust substructures indicated for the HD~100546 disk are approximate, since the location of dust features differs by a few 10s of au along different projections due to the azimuthally asymmetric dust emission in this source \citep{Pineda19, Perez20, Fedele21}. The dip in the emission surface of the HD~100546 disk is coincident with the inner edge of a wide (${\sim}$40-150~au) continuum gap \citep{Pineda19, Fedele21}. In CI~Tau, the vertical dip in CO emitting heights also aligns with a mm dust ring. A similar vertical dip around 50~au is seen in the $^{13}$CO J=3--2 emission surface of this disk as modeled by \citet{Rosotti21}. This is consistent with previous observations showing that vertical substructures often occur at a similar radius in multiple CO isotopologues \citep{Law21}. In DoAr~25, the B86 dust ring \citep{Huang18} lies at the same location as the change in emission surface slope. Similarly, the slope change in CI~Tau is at approximately the same radii as a mm dust ring \citep{Clarke18,Long18}.

All sources with vertical substructure in their emission surfaces also have evidence for kinematic planetary signatures (KPSs). \citet{Pinte20} reported localized deviations from Keplerian rotation, i.e., velocity ``kinks," in the GW~Lup, WaOph~6, and DoAr~25 disks that were inferred directly from individual CO channel maps. Although we do not identify any definitive substructures in the GW~Lup and WaOph~6 disks, both show tentative dips at the same radial locations as the proposed planets. We find no corresponding feature in the CO emission surface of the DoAr~25 disk but note the tentative nature of the KPS in this source \citep{Pinte20}. In the CI~Tau disk, \citet{Rosotti21} identified a similar kinematic signature with a possible planetary origin at 1\farcs3 (${\approx}$210~au). However, this feature is close to the maximum radius at at which we could constrain the CO emission surface and where the SNR is considerably lower. This results in large vertical scatter beyond ${\approx}$150~au and precludes any conclusions about the presence of vertical substructures at large radii. In this disk, \citet{Clarke18} also proposed that the annular continuum gaps - one of which aligns with the vertical dip at 50~au - are due to three Jupiter-mass planets. Since these inferences were based on dust and gas hydrodynamical simulations, it is possible that the other two gaps are, in fact, planetary in origin but do not produce vertical perturbations in the CO line emission surfaces that are detectable with our current data quality. In the HD~100546 disk, a KPS in the form of a Doppler flip was identified at ${\approx}$0\farcs2-0\farcs3, or ${\approx}$20-30~au \citep{Casassus19, Perez20}. While we find a smoothly varying CO emission surface at these radii, a relatively wide vertical dip is present in the emitting heights a few tens of au exterior to this KPS. The proposed locations of two Jupiter-mass planets, one at 15~au and another at 110~au, from the smoothed-particle-hydrodynamical simulations (\citealt{Fedele21}; but see \citealt{Ackermann21} for alternate predictions of planet radial locations at 13~au and 143~au) are located at the inner and outermost edges, respectively, of where we constrained the CO emission surface. Similar to the KPS in the CI~Tau disk, we are unable to determine if any corresponding vertical substructures are present in HD~100546 at or near these radii.

\begin{figure*}[ht!]
\centering
\includegraphics[width=0.8\linewidth]{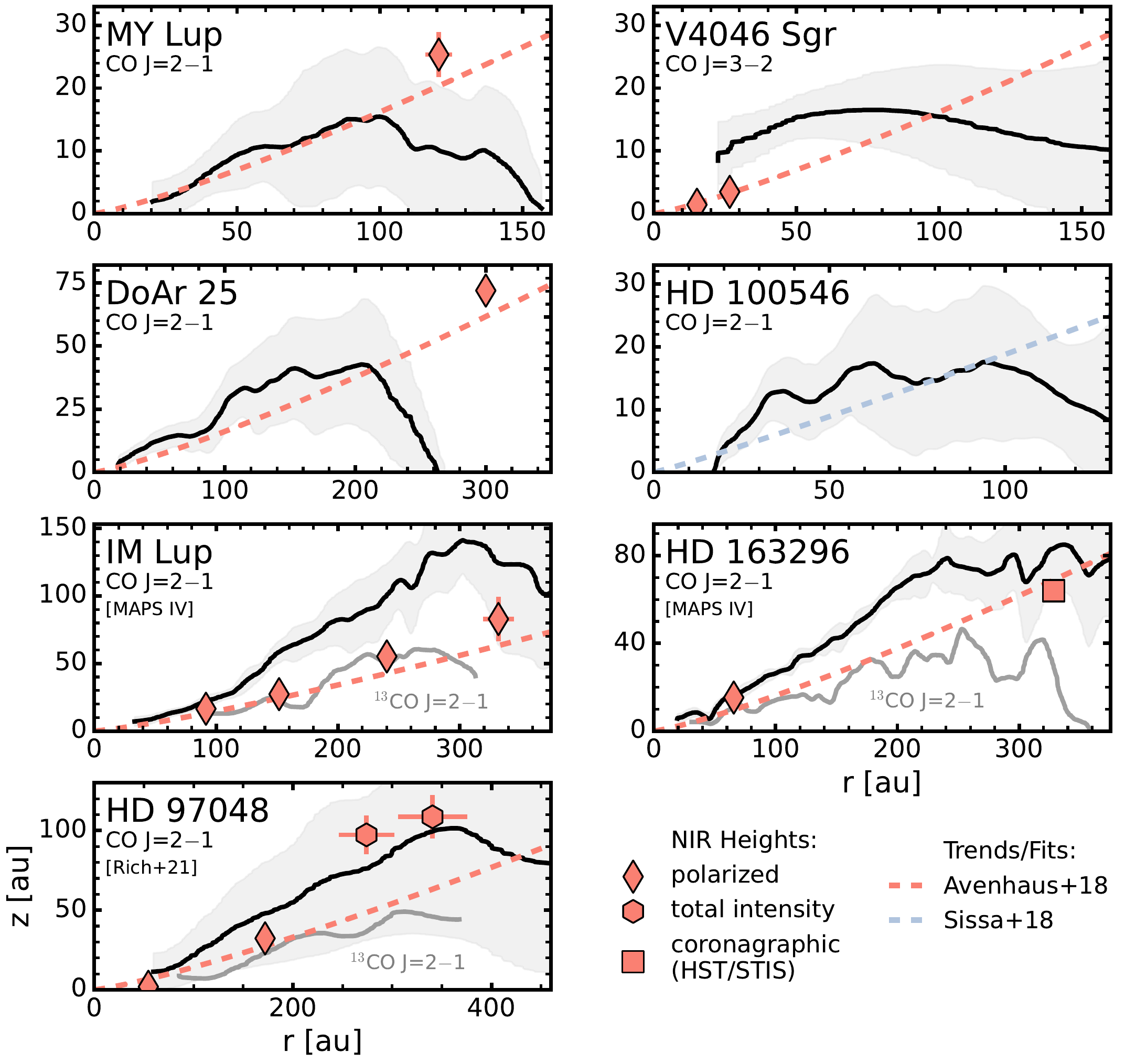}
\caption{CO emission surfaces for sources in our sample (MY~Lup, V4046~Sgr, DoAr~25, HD~100546) and from the literature (HD~163296, IM~Lup, HD~97048) versus NIR heights. The black lines are the moving average surfaces and gray shaded regions show the 1$\sigma$ uncertainty. The red markers show individual height measurements of NIR rings for MY~Lup and V4046~Sgr \citep{Avenhaus18,d_Orazi19}; IM~Lup and HD~163296 \citep{Monnier17, Muro_Arena18, Avenhaus18, Rich20_HST}; and HD~97048 \citep{Ginski16, Rich21}, or from the opening angle at the last scattering separation for DoAr~25 \citep{Garufi20_DARTTS}. Marker types indicate measurements from polarimetric (diamond), total intensity (hexagon), and coronagraphic imaging (square). The red dashed line shows the inferred NIR surface using the powerlaw relation found in a sample of disks in \citet{Avenhaus18}. The blue dashed line shows the geometric scattered light model of the HD~100546 disk from \citet{Sissa18}. The errorbars are smaller than the marker for the rings in V4046~Sgr, while uncertainties are not reported for the NIR measurement in DoAr~25. Light gray curves show the $^{13}$CO J=2--1 emission surfaces in the IM~Lup and HD~163296 \citep{Law21} and HD~97048 disks (Appendix \ref{sec:appendix_HD97048}).}
\label{fig:NIR_compare}
\end{figure*}

\subsection{Comparison with NIR Scattering Surfaces} \label{sec:comparison_NIR_rings}

The vertical distribution of micron-sized dust grains in disks should be related to the gas environment, due to strong coupling between small dust and gas. However, few independent height measurements of both small dust grains and line emission surfaces exist in protoplanetary disks \citep[e.g.,][]{Dutrey17, Villenave20, Rich21, Law21, Flores21, Villenave22} but are critical in probing disk characteristics such as gas-to-dust ratios and turbulence levels. 

Many disks in our sample have been observed in scattered light \citep{Benisty10, Avenhaus14, Garufi16, Avenhaus18, Sissa18, d_Orazi19, Garufi20_DARTTS, Mauco20, Brown_Sevilla21, Garufi21_SPHERE}, which provides information about the distribution of micron-sized dust grains. The MY~Lup and V4046~Sgr disks have well-defined rings in the NIR with direct estimates of scattering heights \citep{Avenhaus18,d_Orazi19}. The high inclination of the DoAr~25 disk also allows for an inference of its NIR surface, despite the absence of NIR substructure in this source \citep{Garufi20_DARTTS}. In addition, a geometric model of the NIR structure of the HD~100546 disk has been constructed by \citet{Sissa18}.

Figure \ref{fig:NIR_compare} shows these NIR heights compared to the CO emission surfaces. To enable a more general comparison, we show the CO emission surfaces versus NIR scattering heights previously reported for the IM~Lup, HD~163296, and HD~97048 disks \citep{Law21, Rich21}. We also plot the powerlaw NIR scattering height relation identified in a sample of disks around T~Tauri stars as part of the DARTTS-S program \citep{Avenhaus18} as a dashed red line in Figure \ref{fig:NIR_compare} for all sources, except HD~100546, where we instead show the \citet{Sissa18} relation. We emphasize that the \citet{Avenhaus18} trend is an average profile meant to illustrate a typical scattered light surface, rather than a detailed fit to each source.

In our sample, the NIR surfaces generally lie either at or below the CO emission surfaces with two exceptions toward larger radii in MY~Lup and DoAr~25. The total size of the NIR disk in DoAr~25 is approximately 100~au greater than that of its CO gas disk (Table \ref{tab:disksize}). The NIR height was only inferred at the outer edge (${\sim}$300~au) of the NIR disk \citep{Garufi20_DARTTS}, but still closely follows the \citet{Avenhaus18} trend and if extrapolated to smaller radii, lies at the same height as CO. A similar result is found for MY~Lup, where the NIR height at ${\approx}$120~au is nearly twice as high as that of CO, but if extrapolated to within 100~au, the surfaces agree nearly exactly. 

The fact that the small dust grain disk size is larger than the CO line emission extent in DoAr~25 is particularly interesting and at first difficult to reconcile. It is possible that this is an observational bias from insufficient line sensitivity, which might have led to a nondetection of low intensity, large radii CO emission in this disk. If, instead, there is truly little-to-no gas at 300~au, it is not clear how small dust grains are lofted to and maintained at such large heights (${\approx}$72~au) without gas pressure support. At this distance, CO may be entirely frozen out, making CO line emission a poor tracer of the gas density at these large radii. The derived temperatures in the outer disk (see Section \ref{sec:gas_temperatures}) are close to those expected for CO freeze-out to occur and in the absence of significant CO non-thermal desorption, might explain these observations. Alternatively, this discrepancy in scattered light and line emission sizes may be an indication of a wind that is entraining the small dust as it leaves the disk. Deeper CO line observations of the DoAr~25 disk are required to confirm its true CO line emission radial extent and the underlying gas density distribution.

In the HD~163296 and IM~Lup disks, \citet{Rich21} and \citet{Law21} found that the CO emission surfaces were considerably more elevated than the NIR heights, with the scattering surfaces typically occupying similar heights as the $^{13}$CO emission surfaces \citep{Law21}. The 330~au ring seen in HST coronagraphic imaging is an exception to this trend, and instead lies at nearly the same height as the CO line emission. In the HD~97048 disk, the CO and NIR surfaces were initially thought to lie at the same height \citep{Rich21}, but after re-deriving the emission surfaces (Appendix \ref{sec:appendix_HD97048}), we find that the NIR surfaces lie closer to the $^{13}$CO emission surfaces, with the caveat that the uncertainties in CO emitting heights are large due to the coarse beam size (${\approx}$0\farcs45). For completeness, we also plot the outer two NIR rings in HD~97048, which were only detected via Angular Differential Imaging \citep{Ginski16}, but were not considered in \citet{Rich21} due to concerns that ADI reduction techniques may alter the shape of continuous objects. The heights of these outer rings are comparable to that of the CO emission surface.

Taken together, our results suggest a greater diversity in CO line emission-to-small-dust heights than previously observed with the caveat that NIR and line emission surfaces are not necessarily tracing the same properties in the outer disk regions. It is nonetheless interesting to note that unlike in the inner disks, the NIR heights at large radii are often either comparable to or larger than the CO line emission heights. Higher spatial resolution CO line observations of disks with known NIR features would enable more robust comparisons between the small dust and line emission heights.

\begin{figure*}[th!]
\centering
\includegraphics[width=\linewidth]{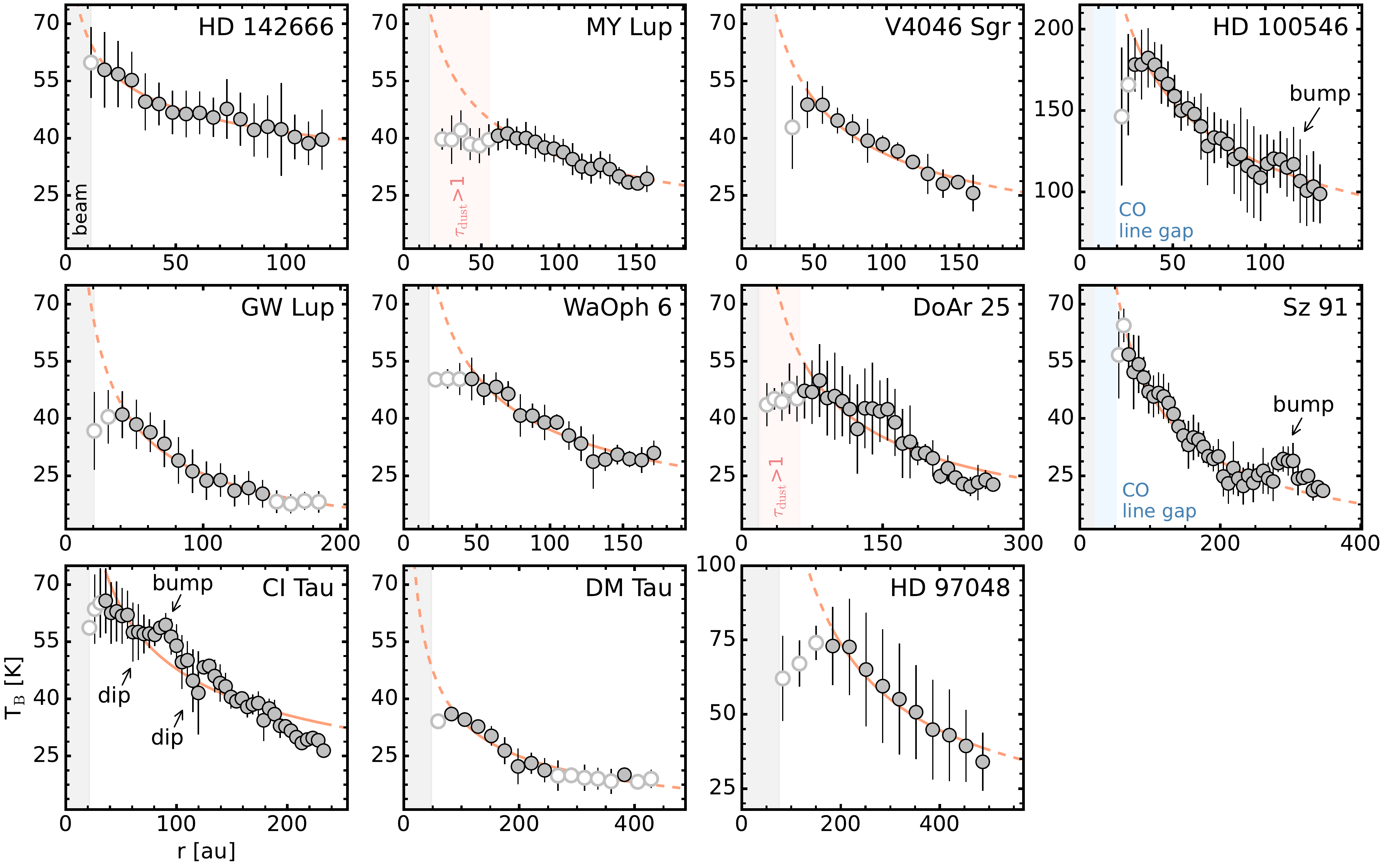}
\caption{CO radial brightness temperature profiles. These profiles represent the mean temperatures computed by radially binning the individual measurements, similar to the procedure used to compute the radially-binned surfaces (see Section \ref{sec:gas_temperatures}). Vertical lines show the 1$\sigma$ uncertainty, given as the standard deviation of the individual measurements in each bin. The solid red lines show the radial range used in the power law fits from Table \ref{tab:radial_temperature_plaw_fits}, while the dashed lines are extrapolations. Temperature measurements affected by dust optical depth, beam dilution, or those below 20~K are marked by hollow markers and are not used in the power law fits. The inner gray shaded region is the FWHM of the beam major axis. Regions of optically thick 1.25~mm continuum emission \citep{Huang18} are shaded in light red in MY~Lup and DoAr~25, while the locations of CO line emission gaps in HD~100546 and Sz~91 (Figure \ref{fig:line_emission_radial_profiles}) are shaded in blue. Temperature bumps are labeled in HD~100546, Sz~91, and CI~Tau with arrows, as are two dips in CI~Tau. All panels show a consistent temperature range, except for the HD~100546 and HD~97048 disks, which are considerably warmer than the other sources.}
\label{fig:figure_temp}
\end{figure*}

\subsection{Gas Temperatures} \label{sec:gas_temperatures}

CO line emission is expected to be optically thick at typical disk temperatures and densities \citep[e.g.,][]{Weaver18}. Assuming the emission fills the beam and is in local thermodynamic equilibrium, the peak surface brightness I$_{\nu}$ provides a measure of the temperature of the emitting gas. Thus, we can use the line brightness temperatures of the extracted emission surfaces to map the disk thermal structure.

\subsubsection{Calculating Gas Temperatures} \label{sec:calc_gas_temperatures}

As a first step, we reran the surface extraction procedure on the line+continuum image cubes to not underestimate the line intensity along lines of sight containing strong continuum emission \citep[e.g.,][]{Boehler17}. For each of the pixels extracted, we obtained a corresponding peak surface brightness and then used the full Planck function to convert I$_{\nu}$ to a brightness temperature, which we assumed is equal to the local gas temperature. We emphasize that all subsequent radial and 2D gas temperature distributions represent those derived directly from these individual surface measurements, i.e., pixels where we were able to determine an emission height.

Several of the disks in our sample suffer from foreground cloud contamination (Table \ref{tab:disk_char}). To avoid underestimating peak brightness temperatures, we manually excluded all channels with cloud obscuration when refitting the line+continuum surfaces. In addition to our sample, we include the HD~97048 disk in the following analysis. While this disk has a previously mapped CO emission surface \citep{Rich21}, it lacks an empirical estimate of its CO temperature structure.

\begin{deluxetable*}{lcccccc}[t!]
\tablecaption{CO Radial Temperature Profile Fits\label{tab:radial_temperature_plaw_fits}}
\tablewidth{0pt}
\tablehead{
\colhead{Source} & \colhead{Line} & \colhead{r$_{\rm{fit, in}}$ [au]}  &\colhead{r$_{\rm{fit, out}}$ [au]} & \colhead{T$_{100}$ [K]} &  \colhead{q} &  \colhead{Feat.\tablenotemark{a}}} 
\startdata
HD~142666 & J=2$-$1 & 18 & 116 & 42~$\pm$~0.5 & 0.20~$\pm$~0.01 & \\
MY~Lup & J=2$-$1 & 61 & 157 & 35~$\pm$~0.3 & 0.41~$\pm$~0.03 & \\
V4046~Sgr & J=3$-$2 & 45 & 160 & 36~$\pm$~0.7 & 0.49~$\pm$~0.04 & \\
HD~100546 & J=2$-$1 & 30 & 130 & 116~$\pm$~1.4 & 0.42~$\pm$~0.02 & B110\\
GW~Lup & J=2$-$1 & 41 & 143 & 26~$\pm$~0.5 & 0.60~$\pm$~0.04 & \\
WaOph~6 & J=2$-$1 & 47 & 171 & 37~$\pm$~0.6 & 0.47~$\pm$~0.03 & \\
DoAr~25 & J=2$-$1 & 67 & 268 & 44~$\pm$~1.1 & 0.54~$\pm$~0.05 & \\
Sz~91 & J=3$-$2 & 69 & 250 & 47~$\pm$~0.7 & 0.70~$\pm$~0.03 & B300\\
CI~Tau & J=3$-$2 & 36 & 233 & 48~$\pm$~0.7 & 0.42~$\pm$~0.03 & D70,B90,D120\\
DM~Tau & J=2$-$1 & 83 & 382 & 34~$\pm$~1.0 & 0.47~$\pm$~0.05 & \\
HD~97048 & J=2$-$1 & 184 & 487 & 122~$\pm$~7.8 & 0.72~$\pm$~0.06 & \\
\enddata
\tablenotetext{a}{Local temperature bumps (B) or dips (D) labeled according to their approximate radial location in au.}
\end{deluxetable*}

\vspace{-20pt}
\subsubsection{Radial and Vertical Temperature Profiles}

\begin{figure*}[]
\centering
\includegraphics[width=\linewidth]{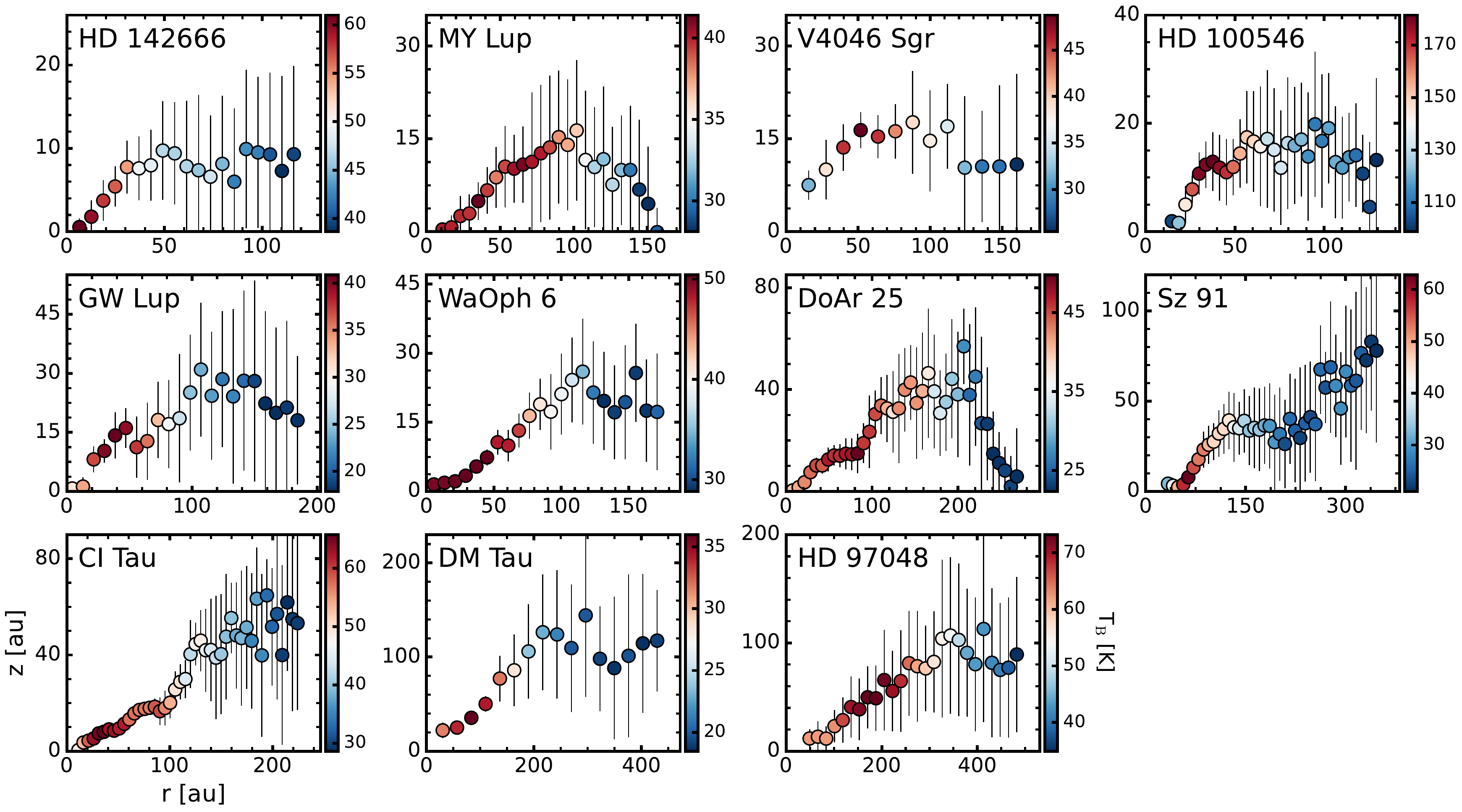}
\caption{2D temperature distributions of CO emission surfaces in all disks. Points are those from the binned surfaces and error bars are the 1$\sigma$ uncertainties in $z$. For some of the innermost points, the uncertainty is smaller than the marker. The uncertainty of the temperature measurements, which is not shown here, can be seen in Figure \ref{fig:figure_temp}. The 2D temperature profiles shown in this figure are available as Data behind the Figure.}
\label{fig:2D_temp_surfaces}
\end{figure*}

Figure \ref{fig:figure_temp} shows the CO radial temperature distributions along the emission surface for all disks. Temperatures range from ${\lesssim}$20~K (DM~Tau) to a maximum of 180~K (HD~100546). Derived brightness temperatures are generally consistent with expectations based on stellar luminosity and spectral classes, with the disks around Herbig Ae/Be stars HD~142666, HD~100546, and HD~97048 showing warmer temperatures than most of the T~Tauri stars. Among the disks around T~Tauri stars, there are modest temperature variations. For instance, the disk around Sz~91 is 1.3-1.5$\times$ warmer than that around DM~Tau at the same radii, despite both being transition disks with similar host stellar luminosities. However, the central cavity of the Sz~91 disk is much larger than that of DM~Tau \citep{Andrews11, Canovas15, Kudo18, Mauco20}, which results in increased irradiation at large radii and likely contributes to this temperature difference. Moreover, we find that the derived temperatures in DM~Tau are consistent with those inferred in the parametric forward models of \citet{Flaherty20}, which account for beam smearing. This suggests that the temperatures derived here are not substantially lowered by non-unity beam filling factors, despite the DM~Tau data having a relatively coarse beam size.

A drop or flattening in brightness temperature is seen interior to 20–50~au in all disks, which is marked as a gray shaded region in Figure \ref{fig:figure_temp}. At the smallest radii, this is primarily due to beam dilution as the emitting area becomes comparable to or smaller than the angular resolution of the observations. However, for the MY~Lup, HD~100546, WaOph~6, and DoAr~25 disks, the central temperature dip or plateau extends further than the beam size. There are several explanations for this: CO is depleted enough for the lines to become optically thin at these radii, the presence of unresolved CO emission substructure, or a substantial fraction of the CO emission is absorbed by dust. The dip in the HD~100546 disk is likely due to the inner CO line emission gap (Figure \ref{fig:figure1}), which results in the emission becoming less optically thick within 1/2-1 beams of the gap edge and thus no longer measures the gas temperature. The inner disks of MY~Lup and DoAr~25 show optically thick dust \citep{Huang18} and the radii where $\tau_{1.25\,\rm{mm}}>$1 are similar to where the derived CO temperature begins to plateau. WaOph~6, however, does not exhibit optically thick dust in its inner disk, but shows hints of additional CO line emission substructure in the form of a low-contrast dip at small radii, as seen in its radial profile in Figure \ref{fig:line_emission_radial_profiles}. Higher angular resolution CO line observations toward this disk are necessary to confirm the reality of this dip and the presence of any additional chemical substructures. 

Next, we fitted the temperature profiles with power law profiles, parameterized by slope $q$ and T$_{100}$, the brightness temperature at 100~au, i.e.,

\begin{equation}
T = T_{100} \times \left(\frac{r}{\rm{100\,au}} \right)^{-q}.
\end{equation}

For derived brightness temperatures less than 20 K -- below the CO freeze-out temperature -- the associated line emission is at least partially optically thin and thus only provides a lower limit on the true gas temperatures. We exclude all temperatures ${<}$20~K in our fits, as well as those affected by beam dilution or dust optical depth, as discussed above (also see Figure \ref{fig:figure_temp}). We also manually excluded the temperature bump at large radii in the Sz~91 disk. We then fitted each profile using the Levenberg-Marquardt minimization implementation in \texttt{scipy.optimize.curve\_fit}. Table \ref{tab:radial_temperature_plaw_fits} lists the fitting ranges and derived parameters. As shown in Figure \ref{fig:figure_temp}, most sources are well fitted by power law profiles and with $q\approx0.4$-$0.6$, while HD~142666 has a considerably shallower ($q=0.20$) profile, and Sz~91 and HD~97048 are steeper ($q=0.70$-$0.72$).

Instead of showing the derived temperature profiles as only a function of radius as in Figure \ref{fig:figure_temp}, we can also map out full 2D temperature profiles. Figure \ref{fig:2D_temp_surfaces} shows the thermal structure of the CO emitting layer as a function of ($r$, $z$) for each source.


\vspace{-1mm}
\subsection{Temperature Substructures} \label{sec:temperature_substr}

\begin{figure*}[]
\centering
\includegraphics[width=\linewidth]{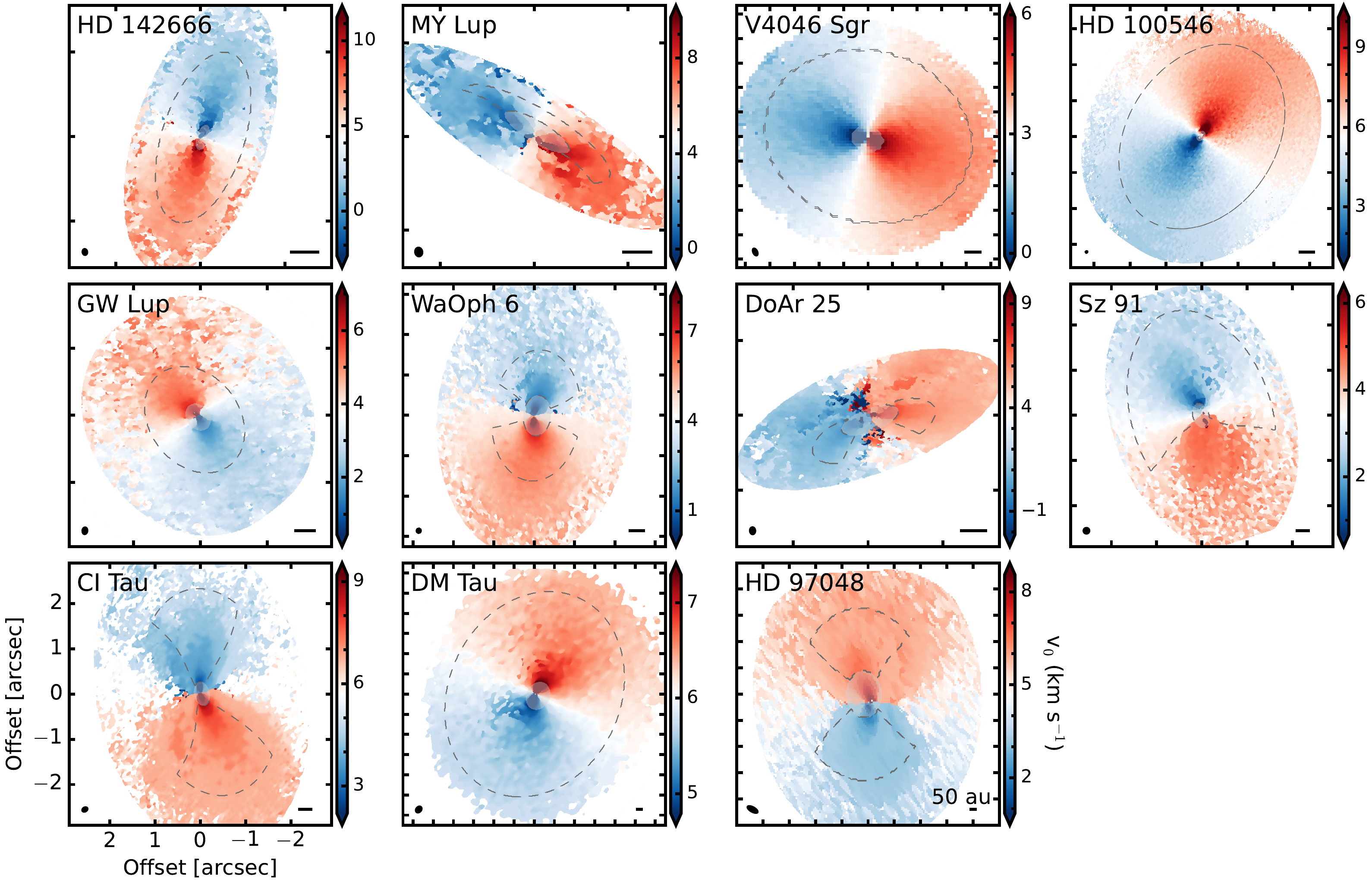}
\caption{Gallery of rotation maps of CO emission in our disk sample. The innermost few beams, which are excluded from the fits, are shaded, while the outermost fitting radius is marked by a dashed line. For those disks with high inclinations or foreground cloud absorption, the wedges used in the fitting are shown by dashed lines. Each tick mark represents 1$^{\prime \prime}$. Velocity signatures from both the front and back sides of the MY~Lup and DoAr~25 disks are clearly visible due to the high inclination of these sources. The synthesized beam and a scale bar indicating 50 au is shown in the lower left and right corner, respectively, of each panel.}
\label{fig:rotation_maps}
\end{figure*}

While the temperature profiles are in general quite smooth, three sources show local dips or bumps in temperature. The HD~100546 and Sz~91 disks show temperature bumps at 110~au and 300~au, respectively, while the CI~Tau disk shows a more complex structure with two dips at 70~au and 120~au and a bump at 90~au. For this 90~au feature in CI~Tau, we are unable to distinguish if this is simply a local maximum resulting from the adjacent dips, or if this is a true temperature enhancement. Each of these features is catalogued in Table \ref{tab:radial_temperature_plaw_fits}.

For these three sources, we checked for possible spatial links with known millimeter dust features, as local temperature deviations in disks are sometimes found at the locations of dust rings or gaps \citep[e.g.,][]{Facchini18, vanderMarel18, Calahan21}.

In HD~100546, the 110~au temperature bump is located at the center of a wide dust gap between the bright inner ring (20-40~au) and the faint outer ring (150-250~au) \citep{Walsh14, Ackermann21}. Recent modeling suggests a 8.5~M$_{\rm{Jup}}$ planet at 110~au (see Figure \ref{fig:substr_plot}) and predicts locally diminished gas and mm dust surface densities \citep{Fedele21}. \citet{Ackermann21} instead find evidence of a 3~M$_{\rm{Jup}}$ planet at 143~au, which places the temperature bump interior to, and not radially coincident with, the proposed planet location.

In Sz~91, the temperature bump at 300~au is well beyond the mm dust ring at 90~au \citep{Canovas16, Mauco20} and corresponds to the low-intensity, plateau-like CO emission seen at large radii (Figure \ref{fig:line_emission_radial_profiles}). A similar temperature bump was identified in the outer disk of IM~Lup \citep{Law21} and is thought to be the result of a midplane temperature inversion \citep{Cleeves16, Facchini17} or due to a photoevaporative wind \citep{Haworth17}.

In CI~Tau, the dip at 120~au aligns with a dust gap, while the dip at 70~au lies close to a $^{13}$CO line emission gap and continuum ring \citep{Long18, Clarke18, Rosotti21}. The 90~au temperature bump is coincident with a pronounced change in the emission surface slope and also close to mm dust ring.

\subsection{Dynamical Masses} \label{sec:appendix_vkep}

We used CO rotation maps to derive dynamical masses for all sources in our sample, closely following the methods of \citet{Teague21}. We first used the `quadratic' method of \texttt{bettermoments} \citep{Teague18_bettermoments} to produce maps of the line center (v$_0$), which includes a statistical uncertainty for v$_0$. The rotation maps were then masked to only include regions where the peak intensities are greater than five times the RMS value measured in a line free channel to remove noisy values at the disk outer edges. 

We fitted the resulting rotation maps with \texttt{eddy} \citep{Teague19eddy}, which uses the \texttt{emcee} \citep{Foreman_Mackey13} python code for MCMC fitting. We consider five free parameters in modeling the Keplerian velocity fields: the source offset from phase center ($\delta x$, $\delta y$), disk position angle (PA), host star mass (M$_*$), and systemic velocity (v$_{\rm{lsr}}$). The disk inclination ($i$) and emission surfaces, parameterized by $z_0$, $\phi$, $r_{\rm{taper}}$, and $\psi$ (Equation \ref{eqn:exp_taper}), were held fixed. For each disk, the innermost 2-4 beams, depending on the source, were masked to avoid confusion from beam dilution. The outermost radii were set by a combination of SNR and the desire to avoid contamination from the rear side of the disk. Table \ref{tab:eddy} provides the selected values. The uncertainty maps produced by \texttt{bettermoments} were adopted as the uncertainties during the fitting. \vspace{-20pt}

\begin{deluxetable*}{llcccccccccccc}[!ht]
\tabletypesize{\scriptsize}
\rotate
\tablecaption{Best Fit CO v$_{\rm{kep}}$ Models \label{tab:eddy}}
\tablewidth{0pt}
\tablehead{
\multicolumn2c{Model} & \colhead{HD 142666} & \colhead{MY Lup\tablenotemark{a}} &  \colhead{V4046 Sgr} & \colhead{HD 100546} & \colhead{GW Lup} & \colhead{WaOph 6\tablenotemark{b}} & \colhead{DoAr 25\tablenotemark{a},\tablenotemark{b}} & \colhead{Sz 91\tablenotemark{b}} & \colhead{CI Tau\tablenotemark{b}} & \colhead{DM Tau} & \colhead{HD 97048\tablenotemark{b}}\vspace{-0.1cm}\\
\multicolumn2c{Parameter} & \colhead{J=2$-$1} & \colhead{J=2$-$1} & \colhead{J=3$-$2} & \colhead{J=2$-$1} & \colhead{J=2$-$1} & \colhead{J=2$-$1} & \colhead{J=2$-$1} & \colhead{J=3$-$2} & \colhead{J=3$-$2} & \colhead{J=2$-$1} & \colhead{J=2$-$1}
}
\rotate \startdata
$\delta x_0$ & (mas) & $-$49 $\pm$ 2 & $-$106 $\pm$ 3 & $-$84 $\pm$ 5 & $-$13 $\pm$ 1 & $-$29 $\pm$ 5 & $-$275 $\pm$ 3 & [38]\tablenotemark{c} & $-$443 $\pm$ 4 & $-$4 $\pm$ 1 & 19 $\pm$ 8 & 19 $\pm$ 7 & \\
$\delta y_0$ & (mas) & 38 $\pm$ 3 & 90 $\pm$ 2 & $-$974 $\pm$ 5 & $-$6 $\pm$ 1 & 6 $\pm$ 5 & $-$341 $\pm$ 5 & [$-$494]\tablenotemark{c} & $-$872 $\pm$ 4 & 9 $\pm$ 1 & $-$21 $\pm$ 10 & 378 $\pm$ 15 & \\
$i$ & ($^{\circ}$) & [$-$62.2] & [$-$73.2] & [34.7] & [$-$41.7] & [$-$38.7] & [$-$47.3] & [67.4] & [49.7] & [$-$49.2] & [36.0] & [$-$41.0] & \\
PA & ($^{\circ}$) & 161.2 $\pm$ 0.29 & 238.4 $\pm$ 0.17 & 255.6 $\pm$ 0.13 & 323.9 $\pm$ 0.05 & 37.2 $\pm$ 0.65 & 173.5 $\pm$ 0.30 & 289.2 $\pm$ 0.36 & 197.0 $\pm$ 0.26 & 192.7 $\pm$ 0.07 & 334.5 $\pm$ 0.24 & 8.0 $\pm$ 0.23 & \\
M$_*$ & (M$_{\odot}$) & 1.73 $\pm$ 0.019 & 1.27 $\pm$ 0.014 & 1.72 $\pm$ 0.008 & 2.10 $\pm$ 0.004 & 0.62 $\pm$ 0.010 & 1.12 $\pm$ 0.008 & 1.06 $\pm$ 0.013 & 0.55 $\pm$ 0.007 & 1.02 $\pm$ 0.001 & 0.50 $\pm$ 0.004 & 2.70 $\pm$ 0.015 & \\
$v_{\rm{LSR}}$ & (km s$^{-1}$) & 4.37 $\pm$ 0.015 & 4.71 $\pm$ 0.014 & 2.93 $\pm$ 0.003 & 5.65 $\pm$ 0.001 & 3.69 $\pm$ 0.011 & 4.21 $\pm$ 0.006 & 3.38 $\pm$ 0.018 & 3.42 $\pm$ 0.005 & 5.70 $\pm$ 0.002 & 6.04 $\pm$ 0.002 & 4.55 $\pm$ 0.004 & \\
$z_0$ & ($^{\prime\prime}$) & [0.09] & [0.21] & [0.28] & [0.35] & [0.22] & [0.37] & [0.31] & [0.91] & [0.32] & [0.82] & [0.88] & \\
$\phi$ & (-) & [0.50] & [1.28] & [0.59] & [1.09] & [0.76] & [1.77] & [1.54] & [2.59] & [1.48] & [1.85] & [2.86] & \\
$r_{\rm{taper}}$ & ($^{\prime\prime}$) & [1.13] & [0.80] & [1.99] & [1.02] & [1.22] & [1.13] & [1.61] & [0.86] & [2.07] & [1.79] & [0.86] & \\
$\psi$ & (-) & [2.37] & [3.95] & [2.59] & [2.57] & [5.91] & [2.52] & [5.85] & [1.99] & [2.61] & [1.67] & [1.10] & \\
$d$ & (pc) & [145.4] & [156.7] & [71.3] & [108.0] & [154.1] & [122.4] & [137.7] & [157.9] & [160.2] & [143.1] & [183.9] & \\
$r_{\rm{fit},in}$ & ($^{\prime\prime}$) & [0.15] & [0.40] & [0.65] & [0.15] & [0.22] & [0.50] & [0.40] & [0.28] & [0.27] & [0.72] & [0.90] & \\
$r_{\rm{fit},out}$ & ($^{\prime\prime}$) & [1.05] & [0.91] & [4.28] & [2.77] & [0.86] & [1.62] & [0.86] & [2.38] & [2.31] & [5.26] & [3.27] & \\
\enddata
\tablecomments{Uncertainties represent the 16th to 84th percentiles of the posterior distribution. Values in brackets were held fixed during fitting.}
\tablenotetext{a}{Due to high disk inclinations, fits performed using manually-drawn wedges to avoid including the back side of the disk.}
\tablenotetext{b}{Wedge sizes and fitting radii were manually adjusted to avoid cloud obscured regions.}
\tablenotetext{c}{R.A. and Dec. positional offsets fixed to those derived from continuum fitting \citep{Huang18}.}
\end{deluxetable*}

We used 64 walkers to explore the posterior distributions of the free parameters, which take 500 steps to burn in and an additional 500 steps to sample the posterior distribution function. The posterior distributions were approximately Gaussian for all parameters with minimal covariance between other parameters. Thus, we took model parameters as the 50th percentiles, and the 16th to 84th percentile range as the statistical uncertainties. Table \ref{tab:eddy} lists the fitted values and uncertainties for all disks.

For disks with foreground cloud absorption, we restricted the fitting regions by using manually selected wedges. The high inclination of MY~Lup and DoAr~25 results in the presence of conspicuous velocity signatures from the back side of the disk. To avoid confusion in the fitting, we also excluded these regions in both disks. Figure \ref{fig:rotation_maps} shows all rotation maps and the fitting regions used in \texttt{eddy}.

\begin{figure}[]
\centering
\includegraphics[width=\linewidth]{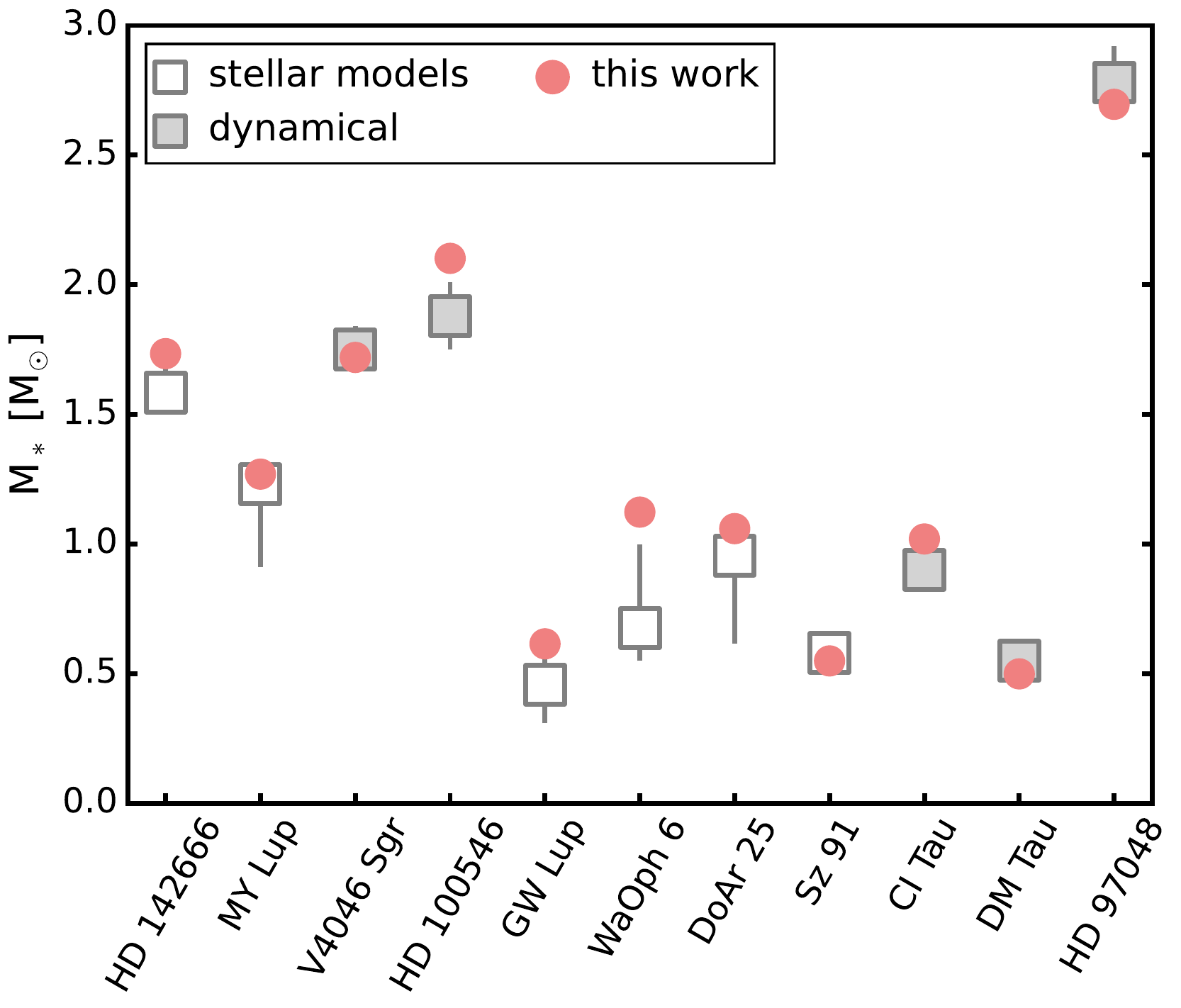}
\caption{Comparison of literature stellar masses (squares) and those derived in this work (circles). Dynamical masses are shown as filled markers while those from stellar models are hollow. The errorbars are smaller than the marker for several sources. The references for literature masses are: HD~142666, MY~Lup, GW~Lup, WaOph~6, DoAr~25 \citep{Andrews18}; V4046~Sgr \citep{Rosenfeld12}; HD~100546, HD~97048 \citep{Casassus19}; Sz~91 \citep{Mauco20}; CI~Tau \citep{Simon17, Simon19}; and DM~Tau \citep{Simon19}.}
\label{fig:compare_masses_plot}
\end{figure}

Figure \ref{fig:compare_masses_plot} shows the derived dynamical masses versus literature values, compiled from both dynamical- and stellar evolutionary model-based estimates. In general, we find excellent agreement with previous measurements, with the exception of WaOph~6, where we find a considerably larger mass (${\approx}$1.1-2.0$\times$) than reported in \citet{Andrews18}. This difference may reflect the uncertainty of stellar evolutionary models in inferring the masses of low-mass pre-main-sequence stars \citep[e.g.,][]{Simon19, Pegues21}, or alternatively, indicate that the spectral type is underestimated by 1-2 subclasses, i.e., WaOph~6 may be a K4/K5-type star instead of K6.

\vspace{-6pt}
\section{Discussion} \label{sec:discussion}
\subsection{Comparison with Previous Results}

The CO emission surfaces of three of our disks have been presented in previous publications using several different methods but with the same data sets as in this work. It is therefore useful to compare their results with ours.

\subsubsection{HD~100546}
\citet{Casassus19} found a CO emission height\footnote{The authors fitted the opening angle $\psi = \arctan (z/r)$ above the disk midplane and found $\psi= 9^{\circ}.3 \pm 2^{\circ}.5$, which is equal to $z$/$r\approx0.16\pm0.04$.} of $z/r{\approx}0.16$ between 0\farcs15-0\farcs75 (17-83~au) by fitting the CO J=2--1 rotation map, i.e., using deviations from Keplerian velocity to infer an emission surface. In this same region, we find $z$/$r\approx0.25$-$0.3$, a factor of two greater than their estimate. We can think of two possible explanations for this discrepancy: (1) The surface begins to flatten and turnover at 0\farcs60 (${\approx}$65~au), and \citet{Casassus19} may have weighted this part of the disk in their fit more than we did, resulting in an overall lower $z/r$, i.e., at 0\farcs75 we find $z/r{\approx}$0.18. (2) We identify a vertical dip at 45~au (Section \ref{sec:vertical_substr_vs_mm_cont}) in the emission surface, which will lower the average $z/r$.

\subsubsection{CI~Tau}
\citet{Rosotti21} found $z/r \approx 0.3$ for the CO J=3--2 emission height, which was visually determined by overlaying conical surfaces onto moment maps of CI~Tau. Overall, this is quite consistent with what we derive, with the caveat that we find a flaring surface such that interior to 90~au, the slope is shallower with $z/r\approx 0.2$-$0.25$, while beyond 90~au, it is $z/r \approx 0.3$.

\subsubsection{DM~Tau}
\citet{Flaherty20} modeled CO line observations in DM~Tau and extracted the resulting CO J=2--1 line emission heights (see their Figure 2). In the inner, flared region of the surface, \citet{Flaherty20} estimated $z/r\sim0.4$, while we found $z/r \gtrsim 0.5$. Beyond 250~au, once the surface begins to plateau, both our directly-mapped surfaces and the modeled emission surfaces lie at roughly the same vertical heights. Thus, we find in general, good agreement between the two approaches.

\subsection{Origins of CO Emission Surface Heights} \label{sec:origins_CO_emission_height}
Given the observed diversity in CO emitting heights, we explore possible mechanisms which may set the vertical extent and degree of flaring in line emission surfaces in the following subsections. We examine trends in emission surface heights with physical characteristics of our sources in Section \ref{sec:Correlations_source} and present possible explanations for the observed correlations in Section \ref{sec:toy_model}.



\subsubsection{Correlations with Source Characteristics} \label{sec:Correlations_source}

We expect that source physical characteristics will influence line emission surfaces. As part of MAPS, \citet{Law21} found that protoplanetary disks with lower host star masses, cooler temperatures, and larger CO gas disks had CO emission surfaces with higher $z/r$ values. However, these trends were tentative, given the small sample size of five disks. \citet{Garufi21} also reported a positive trend between disk size and H$_2$CO line emitting heights in five Class~I disks in the ALMA-DOT survey. This suggests that this trend may extend to earlier phases of disk evolution and may hold for other molecules besides CO, but firm conclusions were again limited by the small sample size. To test the robustness of these trends, we combine our disk sample with the five MAPS disks \citep{Law21} and the HD~97048 disk \citep{Rich21}, which both have CO emission surfaces mapped in the same way.

We first require stellar masses, gas temperatures, and CO gas disk sizes for all sources to enable a homogeneous comparison. We derived dynamical masses (Section \ref{sec:appendix_vkep}) and gas temperatures (Section \ref{sec:gas_temperatures}) for the disks in our sample, while the MAPS disks have existing dynamical masses and CO gas temperatures, which were derived in a consistent way from \citet{Teague21} and \citet{Law21}, respectively. We also computed the CO gas sizes (R$_{\rm{CO}}$) of each disk, as defined by the radius which contained 90\% of total line flux \citep[e.g.,][]{Tripathi17, Ansdell18}. This definition is consistent with that used in \citet{LawMAPSIII} and allows us to easily compare with the CO gas disk sizes of the MAPS sources. Table \ref{tab:disksize} shows the resulting CO sizes and Appendix \ref{sec:appendix_disksize_profiles} provides additional details of this calculation.

Each emission surface spans a range of $z/r$ values, e.g., flaring, plateau/turnover, vertical substructures, but for source-to-source comparisons, we wish to determine a characteristic $z/r$. We choose to focus on the inner regions of the disk where CO emission heights are sharply rising and to exclude the outer disks where the emitting surfaces plateau or turnover. We define the characteristic $z/r$ of each CO emission surface as the mean of all $z/r$ values interior to a cutoff radius of $r_{\rm{cutoff}}=0.8\times$r$_{\rm{taper}}$, where r$_{\rm{taper}}$ is the fitted parameter from the exponentially-tapered power law profiles from Table \ref{tab:emission_surf}. We chose 80\% of the fitted r$_{\rm{taper}}$ to ensure that we only included the rising portion of the emission surfaces and visually confirmed that this choice was suitable for all sources (Figure \ref{fig:zr_definition}). As some disks are considerably more flared than others, i.e., $z/r$ changes rapidly with radius, we also computed the 16th to 84th percentile range within these same radii as a proxy of the overall flaring of each disk. We applied this same definition to the MAPS disks \citep{Law21} and the HD~97048 disk \citep{Rich21} to compile consistent characteristic $z/r$ values. For further details and a list of all $z/r$ values, see Appendix \ref{sec:app:definition_of_zr}.

Figure \ref{fig:zr_literature_correlation} shows these representative $z/r$ values as a function of stellar host mass, mean gas temperature, and CO gas disk size. With this larger disk sample, emission surface heights show a weak decline with both host stellar mass and CO gas temperature. These trends show a high degree of scatter but are broadly consistent with the trends previously seen in \citet{Law21}. We return to these in the following subsection. 

We also find that R$_{\rm{CO}}$ and $z/r$ are strongly correlated. To quantify this correlation, we employ the Bayesian linear regression method of \citet{Kelly07} using the \texttt{linmix} python implementation.\footnote{\url{https://github.com/jmeyers314/linmix}} We find a best-fit relation of $z/r=(3.6\pm0.7\times10^{-4})~\rm{R}_{\rm{CO}} + (0.11 \pm 0.03)$ with a 0.06 scatter of the correlation (taken as the standard deviation $\sigma$ of an assumed Gaussian distribution around the mean relation). We find a correlation coefficient of $\hat{\rho}=0.83$ and associated confidence intervals of (0.44, 0.99), which represent the median and 99\% confidence regions, respectively, of the $2.5\times10^{5}$ posterior samples for the regression. Figure \ref{fig:zr_literature_correlation} shows the derived relationship.



\begin{figure*}[]
\centering
\includegraphics[width=\linewidth]{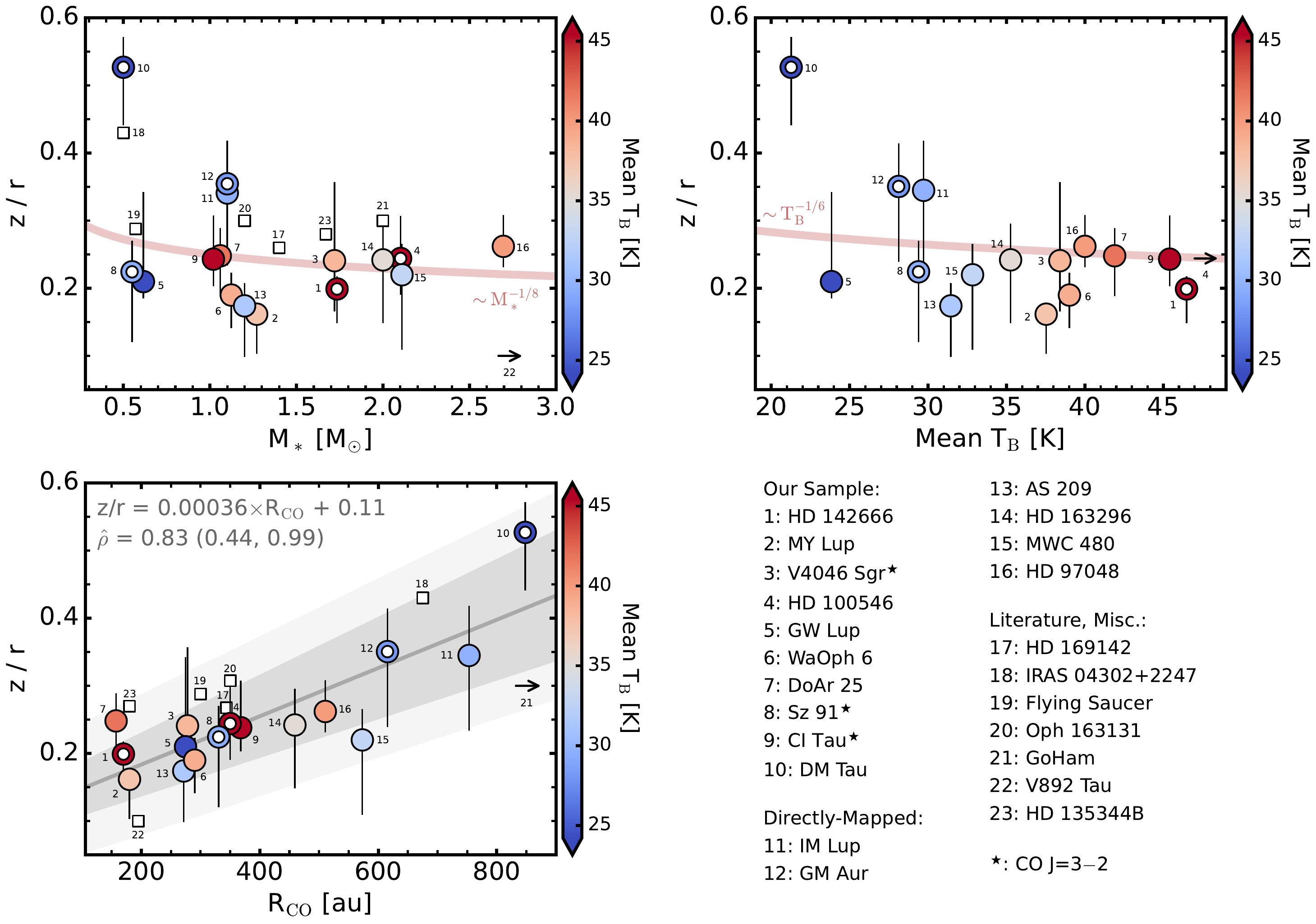}
\caption{Characteristic $z$/$r$ of CO emission heights versus stellar mass (top left), mean CO gas temperature (top right), and CO gas disk size (bottom). All masses are derived dynamically (Section \ref{sec:appendix_vkep}), mean CO gas temperature are computed over the same radial range in which $z$/$r$ is determined, and gas disk sizes are computed as the radius containing 90\% of total flux (Appendix \ref{sec:appendix_disksize_profiles}). Annular markers indicate transition disks. All points are colored by the mean CO gas temperature. Vertical lines show the 16th to 84th percentile range. Approximate scaling relations for stellar mass and temperature are shown as solid red lines. For the $z/r$-R$_{\rm{CO}}$ panel, the derived relation is marked with a solid grey line and the 68\% confidence interval is shown as the dark grey shaded region. The light grey shaded region denotes the scatter around the mean relation. Literature sources without directly-mapped emission surfaces are shown as hollow squares and include: HD~169142 \citep{Fedele17, Yu21}; IRAS~04302+2247 \citep{Podio20}; Flying Saucer \citep{Dutrey17, RR21}; Oph~163131 \citep{Flores21, Villenave22}; GoHam \citep{Teague20_goham}; V892~Tau \citep{Long21}; and HD~135344B \citep{Casassus21}. All data are CO J=2--1, except for those marked with a star ($\star$), which are CO J=3--2.}
\label{fig:zr_literature_correlation}
\end{figure*}

In addition to those sources considered here, we also plot the following literature sources in Figure \ref{fig:zr_literature_correlation} as hollow squares: HD~169142, V892~Tau, HD~135344B (SAO~206462), IRAS~04302+2247, Flying~Saucer (2MASS J16281370-2431391), Oph~163131 (SSTC2D J163131.2-242627), and Gomez’s Hamburger (GoHam, IRAS 18059-3211). HD~169142 is an isolated Herbig Ae/Be star hosting a protoplanetary disk with a CO emission height of $z/r=0.26$ derived from the thermo-chemical models of \citet{Fedele17}. V892~Tau is binary system with two near-equal mass A stars hosting a circumbinary disk with a CO emitting height of $z/r\sim0.1$ inferred directly from channel maps \citep{Long21}, while HD~135344B is an F-type star hosting a transition disk with $z/r= 0.27^{+0.19}_{-0.08}$, as derived from rotation curve fitting \citep{Casassus21}. We measured R$_{\rm{CO}}$ from the radial profiles of HD~169142 \citep{Yu21} and HD~135344B \citep{Casassus21} (see Appendix \ref{sec:appendix_disksize_profiles}), while V892~Tau already had a R$_{\rm{CO}}$ estimate made in a consistent way from \citet{Long21}. The Flying~Saucer \citep{Dutrey17, RuizRodriguez19}, Oph~163131 \citep{Flores21, Villenave22}, and GoHam \citep{Teague20_goham} are edge-on protoplanetary disks, where the emission surface height can be directly measured. IRAS~04302+2247 is an edge-on, Class I disk taken from the ALMA-DOT sample \citep{Garufi21}, with an emission surface of $z/r \approx 0.41$-$0.45$ \citep{Podio20}. For these latter four sources, their edge-on nature makes measuring comparable R$_{\rm{CO}}$ values difficult and we instead visually estimate disk sizes from their zeroth moment maps. The CO gas disk size of IRAS~04302+2247 is particularly uncertain due to presence of envelope emission \citep{Podio20}. All literature sources have existing dynamical mass measurements. Despite their heterogeneous nature, all sources lie closely along the same R$_{\rm{CO}}$-$z/r$ trend as our disk sample. If we include the literature sources in the \texttt{linmix} fitting as before, the derived R$_{\rm{CO}}$-$z/r$ relation remains largely unaltered. Moreover, there do not appear to be any obvious systematic biases affecting emission heights derived from mid-inclination disks versus those inferred directly from edge-on disks.

The GoHam edge-on disk \citep{Teague20_goham} is one notable exception to this trend. The CO emission surface\footnote{This $z/r$ may be modestly underestimated due to the coarse angular resolution (${\gtrsim}1^{\prime \prime}$) of the data from which it was derived \citep{Teague20_goham}. However, this does not change the outlier nature of GoHam, as $z/r$ would need to be more than a factor of two larger to be consistent with the observed $z/r$-R$_{\rm{CO}}$ trend.} is $z/r\sim0.3$, but the size of the CO gas disk is ${>}$1400~au. While one would not necessarily expect the positive R$_{\rm{CO}}$-$z/r$ trend to continue linearly to larger CO gas disks, as this would quickly result in unphysical $z/r$ values, the GoHam value is considerably lower than we see for several other large, e.g., 600-900~au-sized, disks. This suggests that there is some additional effect at play. In the case of GoHam, this lower-than-expected $z/r$ may be due to self-gravity at larger disk radii, especially considering parts of the GoHam disk have been show to be marginally gravitationally unstable, with Toomre parameter $Q\lesssim2$ \citep{Berne15}. It is also possible that GoHam is truly an outlier in terms of its disk structure. Observations of more disks, particularly those with large CO gas extents, are required to assess this.

\subsubsection{Explaining Emission Surface Height Trends} \label{sec:toy_model}

Here, we explore if the trends observed in the previous subsection are in line with expectations based on scaling relations or overall disk structure. In assessing the vertical distribution of line emission in disks, we first consider the gas pressure scale height, $H_{\rm{g}}$, which is given by:
\begin{equation} \label{eqn:Hgas}
    H_{\rm{g}} = \sqrt{ \frac{k_{\rm{B}} T_{\rm{mid}} r^3}{\mu m_p G M_*} }
\end{equation}
where M$_*$ is the stellar mass, $T_{\rm{mid}}$ is the midplane temperature, $k_{\rm{B}}$ is the Boltzmann constant, $\mu$ is the mean molecular weight, $m_p$ is the proton mass, and G is the gravitational constant. For the following discussion, we assume that line emission surface heights correlate with $H_{\rm{g}}$, i.e., $z/r \sim \rm{H}_{\rm{g}}$ and the measured CO gas temperatures in Section \ref{sec:gas_temperatures} correlate with midplane temperature, i.e., $T_{\rm{CO}} \sim T_{\rm{mid}}$. We examine the former assumption in detail in the following subsection and note that while disks have a vertical temperature gradient, as the CO isotopologue data show \citep[][]{Law21}, the perturbations of the vertical structure from an isothermal disk are still generally small \citep[e.g.,][]{Rosenfeld13}. Even if the disk atmosphere temperature traced by CO is substantially warmer than the gas temperature in the midplane, we expect them to at least roughly scale with one another.

\begin{figure}[!h]
\centering
\includegraphics[width=\linewidth]{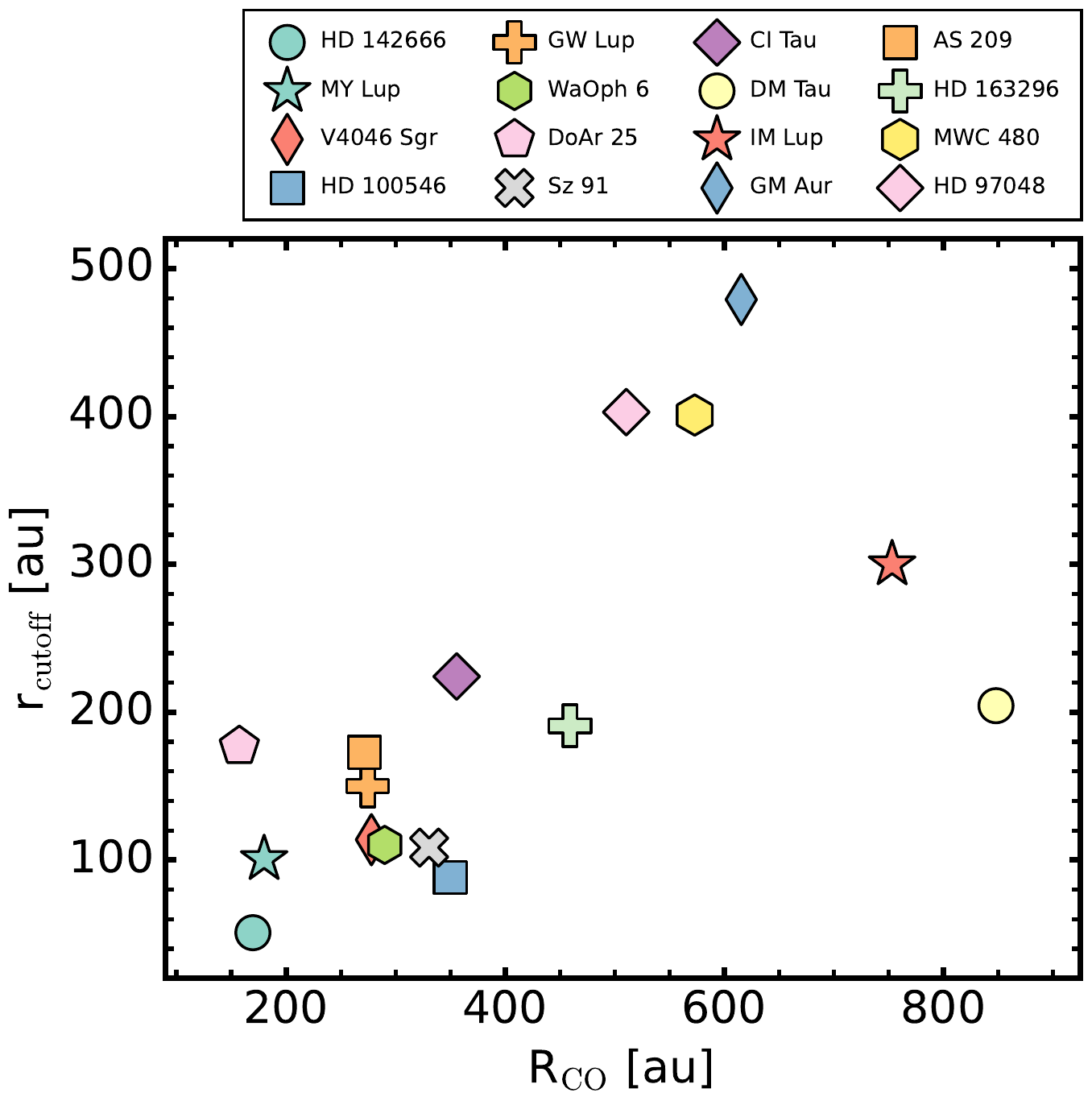}
\caption{CO gas disk size versus the cutoff radius, interior to which characteristic $z$/$r$ values were measured (see Appendix \ref{sec:app:definition_of_zr}).}
\label{fig:rcutoff_vs_zr}
\end{figure}

From Equation \ref{eqn:Hgas}, line emitting heights scale as $z/r \sim \sqrt{T} / \sqrt{M_*}$. Thus, stellar mass and gas temperature should each contribute in setting the $z/r$ of emission surfaces, with cooler disks and less massive host stars leading to more vertically-extended emission surfaces. However, we do not expect T and $M_*$ to be independent variables and to estimate scaling relationships, we next need to examine the expected dependent of T on $M_*$.

If $\rm{T} \sim L_*^{1/4}$, then for any stellar mass-luminosity scaling $L_* \sim M^a$ with $a < 4$, we expect z to weakly decrease with M$_*$. For instance, if $a=3$, then $T \sim M^{3/4}$ and we find that $z/r \sim M^{-1/8}$. If we instead consider temperature instead of stellar mass, we find that $z/r$ also scales weakly with T (again, assuming $a < 4$). As above, for $a=3$, we expect $z/r \sim T^{-1/6}$. Both of these scaling relations are shown in their respective panels in Figure \ref{fig:zr_literature_correlation}. Thus, the weakly declining trends between both $z/r$ and stellar mass and mean CO gas temperature seen in Figure \ref{fig:zr_literature_correlation} are, to first order, consistent with expectations from these simple scaling relations. However, in contrast to the observed $z/r$-R$_{\rm{CO}}$ correlation, these trends remain highly suggestive in nature, especially due to the limited parameter space they span, namely either few or no sources with low (${<}$0.5~M$_{\odot}$) or high (${>}$3~M$_{\odot}$) stellar masses or with warmer (T$_{\rm{B}} > 50$~K) mean gas temperatures.

We next consider the origins of the strong $z/r$-R$_{\rm{CO}}$ correlation observed in the previous subsection. In Figure \ref{fig:rcutoff_vs_zr}, we show the CO gas disk size versus the cutoff radius, in which the characteristic $z/r$ values were measured (also see Appendix \ref{sec:app:definition_of_zr}). We find a positive trend between R$_{\rm{CO}}$ and r$_{\rm{cutoff}}$, which suggests that the $z/r$-R$_{\rm{CO}}$ correlation is due to the flared nature of disk line emission surfaces. As we are averaging over wider radial ranges, i.e., larger r$_{\rm{cutoff}}$, for those disks with larger R$_{\rm{CO}}$, we find higher characteristic $z/r$ values. Thus, we expect the $z/r$-R$_{\rm{CO}}$ trend seen in Figure \ref{fig:zr_literature_correlation} to be driven, in large part, by disk flaring.

\subsection{Emission Surfaces and Gas Scale Heights}

\begin{figure*}[!]
\centering
\includegraphics[width=\linewidth]{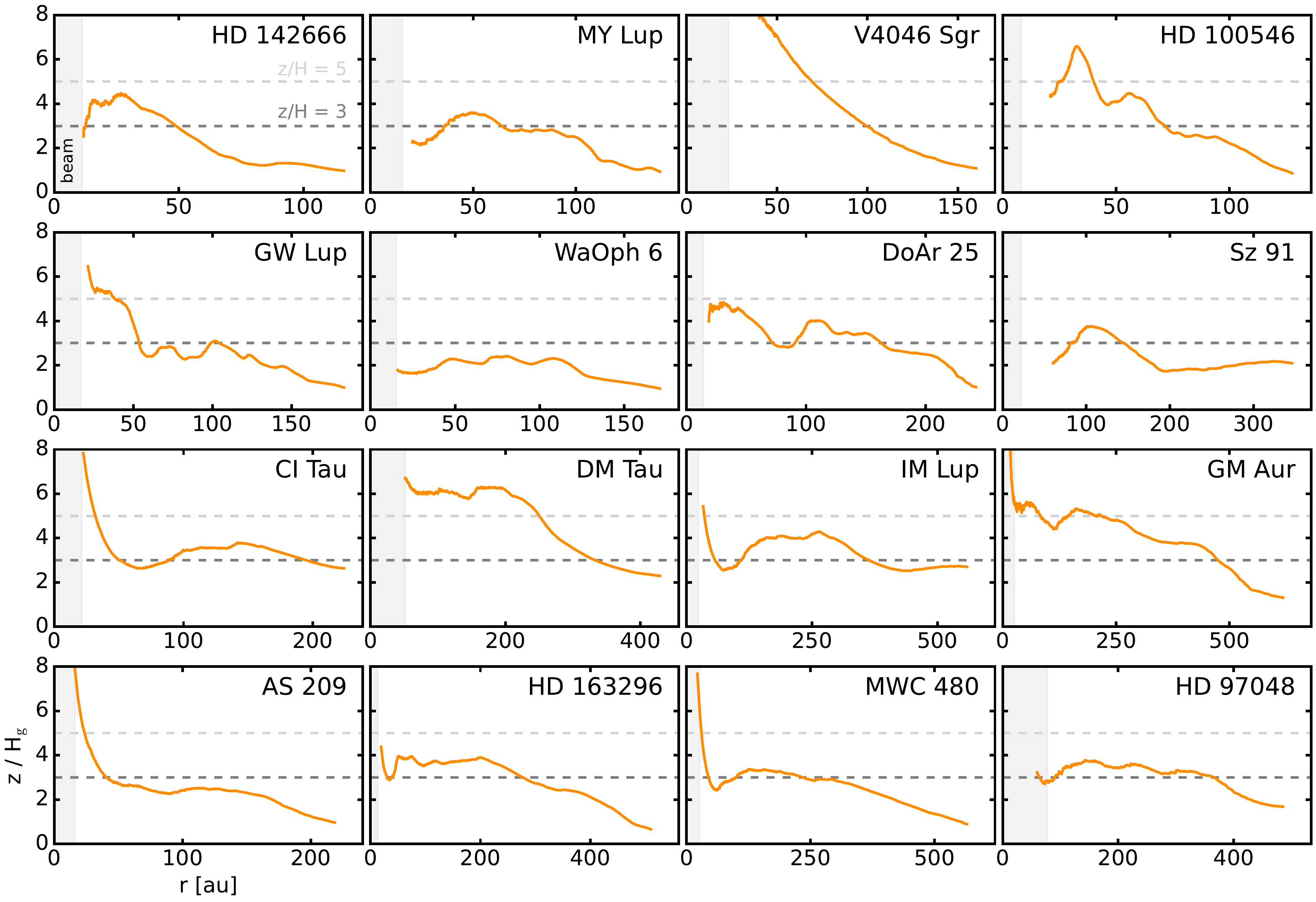}
\caption{Ratio of CO emission surfaces and gas pressure scale heights for all disks in our sample. Dashed gray lines show constant ratios of three and five. Each disk has a different radial range, corresponding to the range where we were able to extract emission surfaces. The inner gray shaded region is the FWHM of the beam major axis.}
\label{fig:Hgas_vs_ZCO}
\end{figure*}

Next, we explore the relationship between CO line emission surfaces and gas pressure scale heights.

We adopt the model of H$_{\rm{gas}}$ from Equation \ref{eqn:Hgas}. We take M$_*$ from Table \ref{tab:disk_char} and assume $\mu=2.37$. We approximate the midplane temperature profile using the simplified expression for a passively heated, flared disk in radiative equilibrium \citep[e.g.,][]{Chiang97,DAlessio98,Dullemond01}:
\begin{equation} \label{eqn:Tmid}
    T_{\rm{mid}} = \left( \frac{\varphi L_*}{8 \pi r^2 \sigma_{\rm{SB}}} \right)^{0.25}
\end{equation}
where L$_*$ is the stellar luminosity (Table \ref{tab:disk_char}), $\sigma_{\rm{SB}}$ is the Stefan-Boltzmann constant, and $\varphi$ is the flaring angle. For consistency with \citet{Huang18} and \citet{Dullemond18}, we adopt a conservative $\varphi=0.02$ for all disks. We note that, if instead, we use the values obtained from our CO emission surfaces and assume that $z/r$ is a perfect tracer of H$_{\rm{g}}/r$, i.e., $\varphi = \psi-1$, we find a constant offset in H$_{\rm{g}}$ by a factor of ${\approx}1.3$, or that the CO line emission surface is more vertically extended than the absorption surface. This is sensible, as disks have vertical temperature inversions and thus more gas at $z = \rm{H}_{\rm{g}}$ - and a higher CO $\tau=1$ emission surface - than expected from a simple Gaussian vertical model.

Figure \ref{fig:Hgas_vs_ZCO} shows the ratio of the CO emission surfaces and derived H$_{\rm{g}}$ as a function of radius, i.e., $z/\rm{H}_{\rm{g}}$. For the majority of disks, the CO emission surface traces ${\approx}$2-5$\times$H$_{\rm{g}}$, which is consistent with previously inferred ratios between CO emitting heights and H$_{\rm{g}}$ \citep[e.g.,][]{Dartois03, Dutrey17, pinte18, Flaherty20}. Some sources show relatively constant ratios over their radial extents, such as WaOph~6 or HD~97048, while others have ratios that vary by up to a factor three, e.g., GW~Lup, DM~Tau. In a few cases in the innermost radii, the ratio reaches very high values ${\gtrsim}$8, but this is the region in which it is the most difficult to extract emission surfaces. Thus, such high inner values should be regarded with caution.

\begin{figure}[]
\centering
\includegraphics[width=0.9\linewidth]{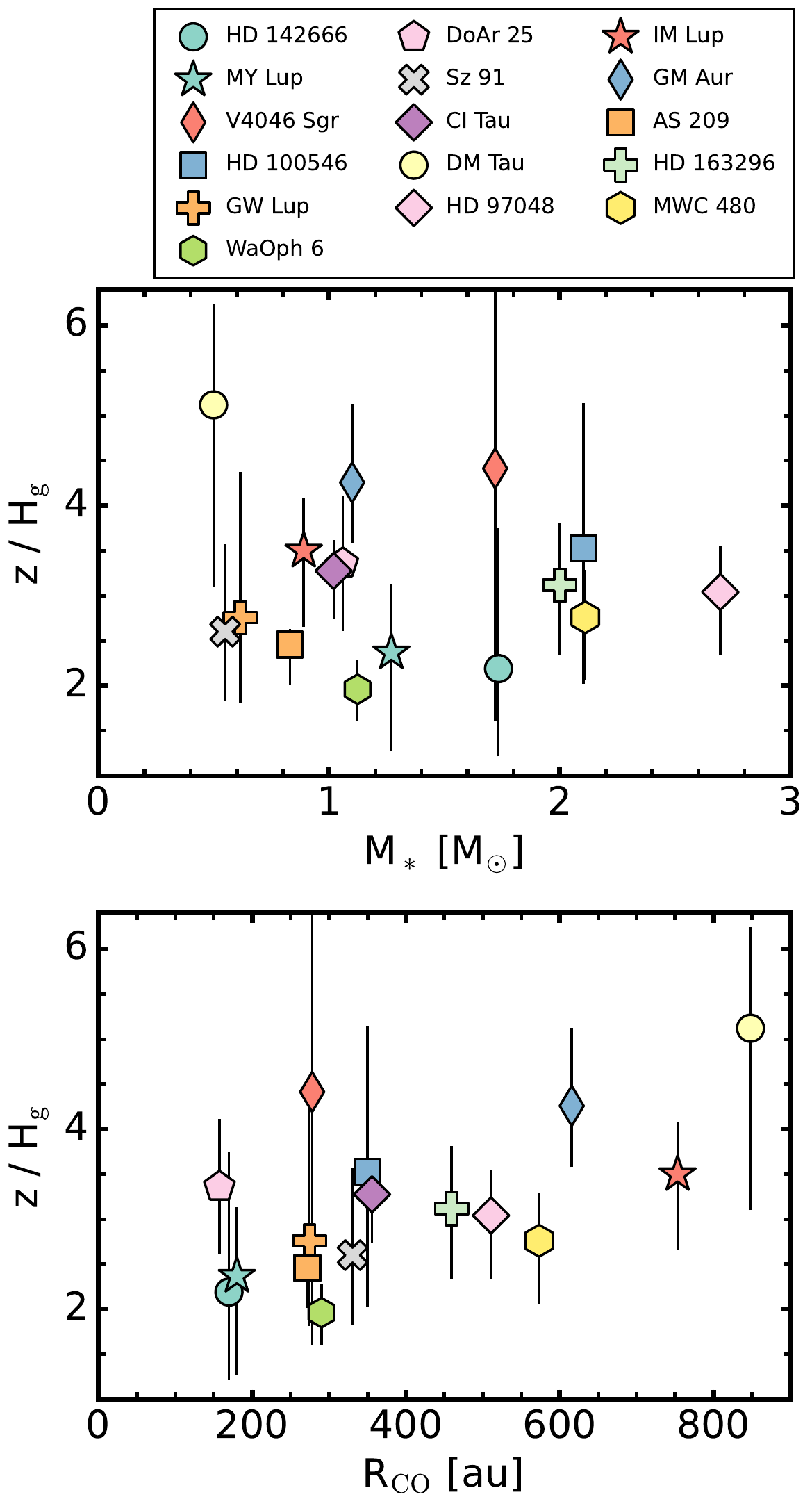}
\caption{Average $z/H_{\rm{g}}$ ratios versus stellar mass (top) and CO gas disk size (bottom). Markers show the 50th percentile, while vertical lines show the 16th to 84th percentile range, calculated from the profiles in Figure \ref{fig:Hgas_vs_ZCO}.}
\label{fig:Hgas_correlation}
\end{figure}

To use CO emission surfaces to infer gas pressure scale heights, we need to better understand why the $z/\rm{H}_{\rm{g}}$ ratios are so different both within and among disks. Here, we explore if the difference can be attributed to stellar mass or disk radius. Figure \ref{fig:Hgas_correlation} shows the mean $z/\rm{H}_{\rm{g}}$ of each disk versus stellar mass and the CO disk gas size. We find that mean $z/\rm{H}_{\rm{g}}$ weakly declines with stellar mass and shows a positive correlation with R$_{\rm{CO}}$. These trends follow those observed in Section \ref{sec:origins_CO_emission_height} but exhibit considerably greater scatter. They are likely driven by the emission surface height correlation seen in Figure \ref{fig:zr_literature_correlation}, as higher emission surfaces will result in larger $z/\rm{H}_{\rm{g}}$ ratios and the fact that source-to-source variation in H$_{\rm{g}}$ does not exceed a factor of two and for most sources, is often considerably smaller. Overall, this suggests that if one can measure both the CO gas disk size and emission surface for a particular disk, it may be possible to infer its radially-averaged gas pressure scale height.

\section{Conclusions} \label{sec:conlcusions}

Using archival ALMA observations of CO J=2--1 and J=3--2 at high spatial resolution, we extracted emission surfaces in a sample of ten protoplanetary disks. We find the following:

\begin{enumerate}
    \item CO line emission surfaces vary substantially among disks in their heights. Peak emission heights span a few tens of au to over 100~au, while $z$/$r$ values range from ${\approx}0.1$ to $\gtrsim$0.5.
    \item A few emission surfaces present substructures in the form of vertical dips or abrupt slope changes. All of these features align with known millimeter dust substructures.
    \item We compare the heights of micron-sized dust grains and CO line emission for those disks with well-constrained NIR scattering heights. CO-to-small-dust heights are quite diverse, with CO emitting heights being higher than the NIR scattering surfaces in some sources, while in others, such as the MY~Lup and DoAr~25 disks, the NIR heights are more elevated than the CO line emission. The radial extent of the DoAr~25 disk in scattered light is nearly 100~au larger than in CO line emission, which may be due to insufficient line sensitivities, the presence of a wind, or CO freeze-out at large radii.
    \item We derive radial and vertical temperature distributions in CO for all disks. Temperatures are generally consistent with source spectral types, and range from ${\lesssim}$20~K in DM~Tau to a peak of 180~K in HD~100546.  A handful of disks show local increases or decreases in gas temperature, some of which correspond to the radial locations of known millimeter dust features or proposed embedded planets.
    \item By combining our sample with literature sources, including the MAPS disks, that have previously mapped CO emission surfaces, we find that emission surface heights weakly decline with stellar host mass and mean gas temperature. Due to the large scatter present, these trends are only suggestive but are generally consistent with expectations from simple scaling relations. We also identify a strong positive correlation between emission surface $z/r$ and CO gas disk size, which is largely due to the flared nature of line emission surfaces in disks.
    \item We compare the derived CO emission surfaces to the gas pressure scale heights in our disk sample. We find that, on average, the CO emission surface traces ${\approx}2$-$5\times$H$_{\rm{g}}$. We also identify a tentative trend between CO gas disk size and the ratio of line emission height and scale height, which suggests that CO line emission surfaces could be calibrated as empirical tracers of average H$_{\rm{g}}$ values.
    \item We also derived dynamical masses and CO gas disk sizes for all disks in our sample. Dynamical masses are consistent with literature estimates, except for WaOph~6 where we find M$_*=1.12$~M$_{\odot}$, which is {$\approx$}1.1-2.0$\times$ larger than previous stellar evolutionary model estimates.
\end{enumerate}

We have shown an effective method for extracting CO emitting layers in a large sample of disks. Such a method can naturally be extended to comparable observations of CO isotopologue lines, which allows a full mapping of 2D disk structure and temperature \citep[e.g.,][]{pinte18, Law21}, or to other important molecular tracers of disk chemistry and structure \citep[e.g.,][]{Teague20,Bergner21ApJS}. Higher sensitivity CO line emission data are also necessary to better characterize the prevalence and nature of vertical substructures, and how they relate to other disk characteristics. \\

This paper makes use of the following ALMA data: ADS/JAO.ALMA\#2012.1.00761.S, 2015.1.00192.S, 2016.1.00315.S, 2016.1.00344.S, 2016.1.00484.L, 2016.1.00724.S, and 2017.A.00014.S. ALMA is a partnership of ESO (representing its member states), NSF (USA) and NINS (Japan), together with NRC (Canada), MOST and ASIAA (Taiwan), and KASI (Republic of Korea), in cooperation with the Republic of Chile. The Joint ALMA Observatory is operated by ESO, AUI/NRAO and NAOJ. The National Radio Astronomy Observatory is a facility of the National Science Foundation operated under cooperative agreement by Associated Universities, Inc.

C.J.L. thanks Gerrit van der Plas, Simon Casassus, Linda Podio, and Christian Flores for providing data for HD~97048, HD~135344B, IRAS~04302+2247, and Oph~163131, respectively.

C.J.L. acknowledges funding from the National Science Foundation Graduate Research Fellowship under Grant No. DGE1745303. R.T. and F.L. acknowledge support from the Smithsonian Institution as a Submillimeter Array (SMA) Fellow. K.I.\"O. acknowledges support from the Simons Foundation (SCOL \#321183) and an NSF AAG Grant (\#1907653). E.A.R acknowledges support from NSF AST 1830728. T.T. is supported by JSPS KAKENHI Grant Numbers JP17K14244 and JP20K04017. S.P. acknowledges support ANID/FONDECYT Regular grant 1191934 and Millennium Nucleus NCN2021080 grant. J.D.I. acknowledges support from the Science and Technology Facilities Council of the United Kingdom (STFC) under ST/T000287/1. S.M.A. and J.H. acknowledge funding support from the National Aeronautics and Space Administration under Grant No. 17-XRP17 2-0012 issued through the Exoplanets Research Program. Support for J.H. was provided by NASA through the NASA Hubble Fellowship grant \#HST-HF2-51460.001-A awarded by the Space Telescope Science Institute, which is operated by the Association of Universities for Research in Astronomy, Inc., for NASA, under contract NAS5-26555. G.P.R. acknowledges support from the Netherlands Organisation for Scientific Research (NWO, program number 016.Veni.192.233). L.M.P. gratefully acknowledges support by the ANID BASAL project FB210003, and by ANID, -- Millennium Science Initiative Program -- NCN19\_171. V.V.G. acknowledges support from FONDECYT Iniciaci\'on 11180904 and ANID project Basal AFB-170002. J.H.K.’s research is supported by NASA Exoplanets Research Program grant 80NSSC19K0292 to Rochester Institute of Technology. J.B. acknowledges support by NASA through the NASA Hubble Fellowship grant \#HST-HF2-51427.001-A awarded by the Space Telescope Science Institute, which is operated by the Association of Universities for Research in Astronomy, Incorporated, under NASA contract NAS5-26555.

%

\facilities{ALMA}


\software{Astropy \citep{astropy_2013,astropy_2018}, \texttt{bettermoments} \citep{Teague18_bettermoments}, CASA \citep{McMullin_etal_2007}, \texttt{disksurf} \citep{disksurf_Teague}, \texttt{GoFish} \citep{Teague19JOSS}, \texttt{keplerian\_mask} \citep{rich_teague_2020_4321137}, \texttt{linmix} (\url{https://github.com/jmeyers314/linmix}), Matplotlib \citep{Hunter07}, NumPy \citep{vanderWalt_etal_2011}, SciPy \citep{Virtanen_etal_2020}}




\appendix

\section{CO zeroth moment maps, radial profiles, and gas disk sizes} \label{sec:appendix_disksize_profiles}

All zeroth moment maps shown in Figure \ref{fig:figure1} were generated using the \texttt{bettermoments} \citep{Teague18_bettermoments} python package, closely following the procedures outlined in \citet{LawMAPSIII}. Briefly, we adopted Keplerian masks generated using the \texttt{keplerian\_mask} \citep{rich_teague_2020_4321137} code and based on the stellar+disk parameters listed in Table \ref{tab:disk_char}. Each mask was visually inspected to ensure that it contained all emission present in the channel maps and if required, manual adjustments to mask parameters were made, e.g., maximum radius, beam convolution size. For accurate flux recovery, we did not use a flux threshold for pixel inclusion, i.e., sigma clipping. Channels containing either no emission or significant absorption due to cloud contamination were excluded. \vspace{-12pt}

\begin{deluxetable}{llcc}
\tablecaption{Gas Disk Sizes \label{tab:disksize}}
\tabletypesize{\normalsize} \tablewidth{0pt}
\tablehead{
\colhead{Source} & \colhead{Line} & \colhead{R$_{\rm{CO}}$} & \colhead{R$_{\rm{edge}}$} \vspace{-0.1cm} \\
\colhead{} & \colhead{CO} & \colhead{[au]} & \colhead{[au]}
}
\startdata
HD~142666 & J=2$-$1 & 170 $\pm$ 4 & 209 $\pm$ 15 \\
MY~Lup & J=2$-$1 & 180 $\pm$ 5 & 231 $\pm$ 14 \\
V4046~Sgr & J=3$-$2 & 278 $\pm$ 7 & 360 $\pm$ 7 \\
HD~100546 & J=2$-$1 & 350 $\pm$ 3 & 480 $\pm$ 4 \\
GW~Lup & J=2$-$1 & 275 $\pm$ 27 & 424 $\pm$ 36 \\
WaOph~6 & J=2$-$1 & 290 $\pm$ 6 & 435 $\pm$ 24 \\
DoAr~25 & J=2$-$1 & 157 $\pm$ 4 & 214 $\pm$ 12 \\
Sz~91 & J=3$-$2 & 331 $\pm$ 6 & 418 $\pm$ 12 \\
CI~Tau & J=3$-$2 & 356 $\pm$ 7 & 571 $\pm$ 19 \\
DM~Tau & J=2$-$1 & 848 $\pm$ 14 & 1055 $\pm$ 23 \\
\textit{HD~97048\tablenotemark{a}} & J=2$-$1 & 511 $\pm$ 21 & 733 $\pm$ 26 \\
\textit{HD~169142\tablenotemark{b}} & J=2$-$1 & 344 $\pm$ 6 & 424 $\pm$ 18 \\
\textit{HD~135344B\tablenotemark{c}} & J=2$-$1 & 180 $\pm$ 31 & 235 $\pm$ 34 \\
\enddata
\tablenotetext{a}{Fit using the radial profile derived from reimaged CO J=2--1 data (see Appendix \ref{sec:appendix_HD97048}).}
\tablenotetext{b}{Fit using azimuthally-averaged radial profile from \citet{Yu21}.}
\tablenotetext{c}{Fit using the azimuthally-averaged radial profile generated from the \textit{uv}-tapered, single Gaussian fit map from \citet{Casassus21}.}
\tablecomments{Disk size (R$_{\rm{CO}}$) and outer edge (R$_{\rm{edge}}$) were computed as the radius which encloses 90\% and 99\% of the total disk flux, respectively.}
\end{deluxetable}

\begin{figure*}[h!]
\centering
\includegraphics[width=\linewidth]{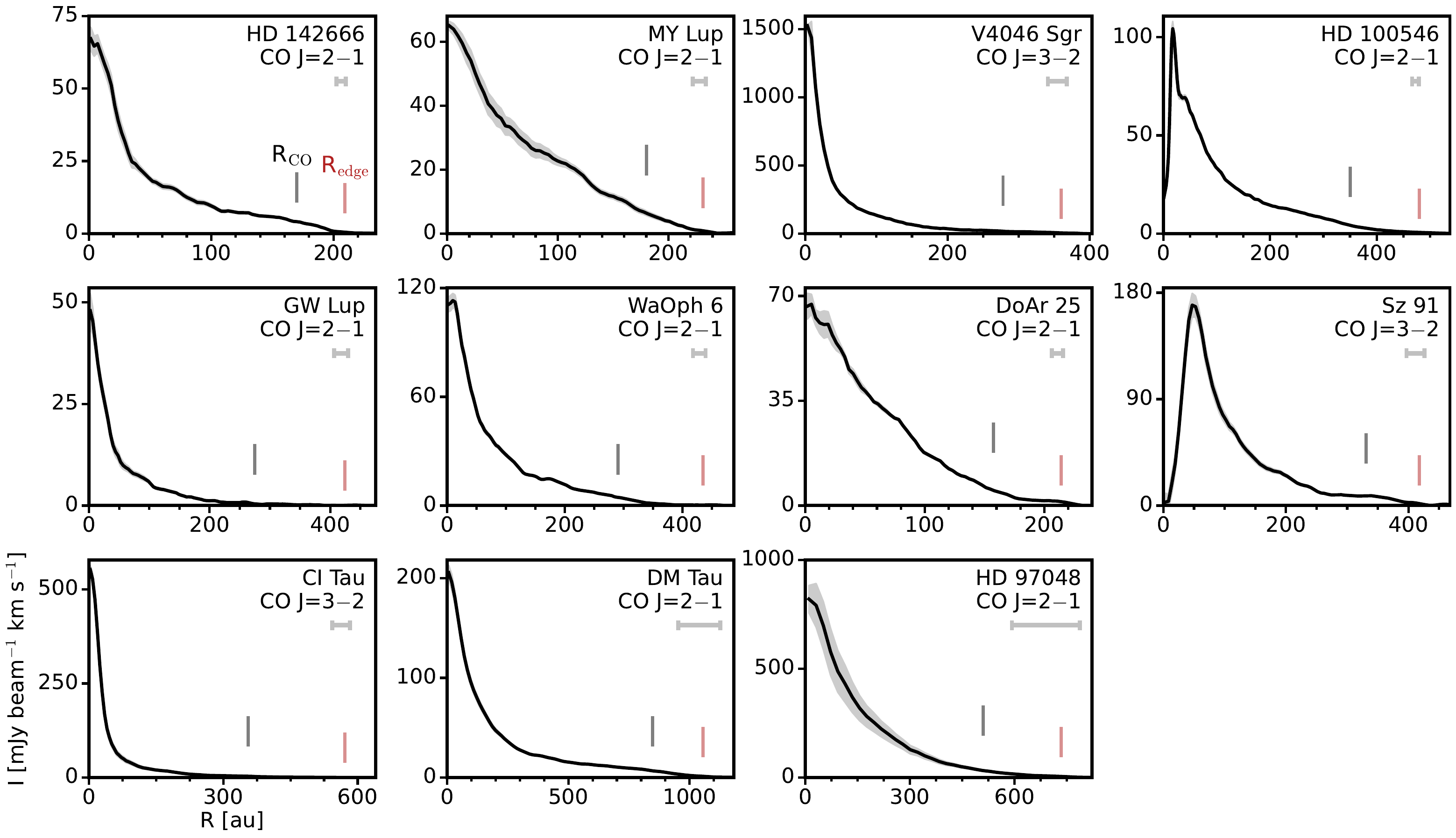}
\caption{Deprojected radial intensity profiles of CO lines for our sample and the HD~97048 disk. Gray shaded regions show the 1$\sigma$ uncertainty, measured as the standard error on the mean in the annulus or arc over which the emission was averaged. The radial locations of R$_{\rm{CO}}$ and R$_{\rm{edge}}$ from Table \ref{tab:disksize} are labeled in gray and red, respectively. The FWHM of the major axis of the synthesized beam is shown by a horizontal bar in the upper right corner of each panel.}
\label{fig:line_emission_radial_profiles}
\end{figure*}

\begin{figure*}[h!]
\centering
\includegraphics[width=\linewidth]{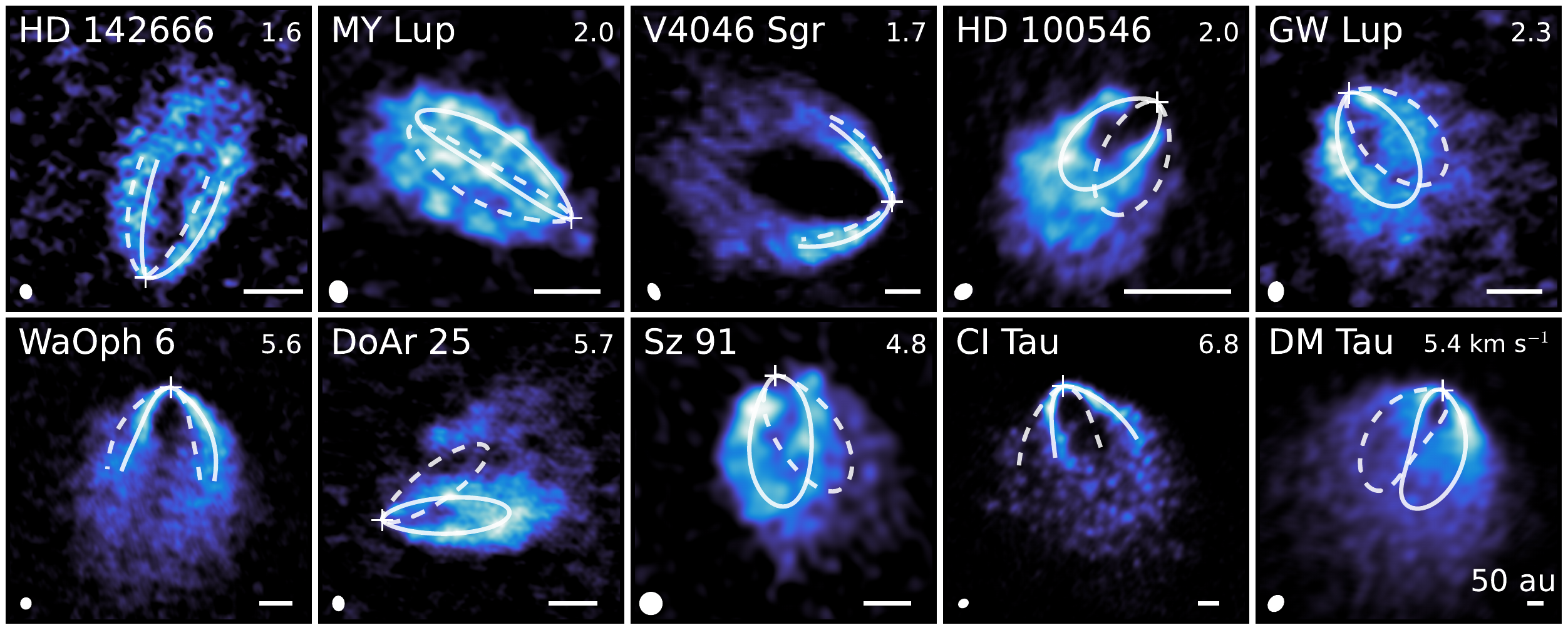}
\caption{Representative CO line emission channels for each of the disks in our sample. The CO isovelocity contours are derived using the parametric fits in Table \ref{tab:emission_surf} and source parameters from Table \ref{tab:eddy}. The extent of the contours corresponds to only those radial regions where we have direct constraints on the line emission surface. Crosses mark the centers of each disk. Solid curves indicate the upper surface of the disk and dashed curves mark the lower surface. Kinematic Local Standard of Rest (LSRK) velocities are marked in the upper right corner. The synthesized beam and a scale bar indicating 50~au is shown in the lower left and right corner, respectively, of each panel.}
\label{fig:isovel_contours}
\end{figure*}

\begin{figure*}[t!]
\centering
\includegraphics[width=0.75\linewidth]{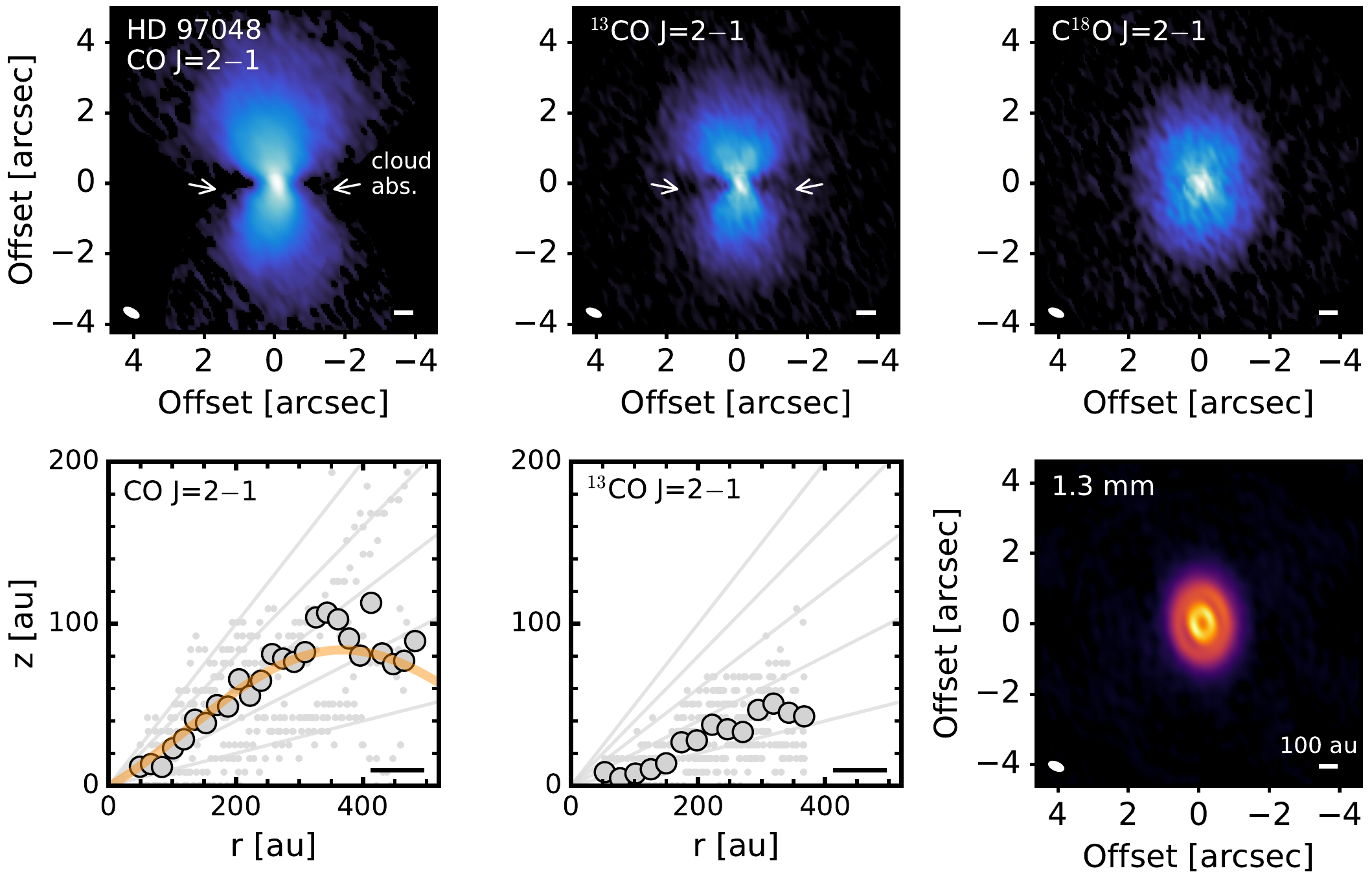}
\caption{From left to right, top to bottom: CO, $^{13}$CO, and C$^{18}$O J=2--1 zeroth moment maps; CO and $^{13}$CO line emission surfaces; and 1.3~mm continuum image. The synthesized beam and a scale bar indicating 100~au is shown in the lower left and right corner, respectively, of each image. Regions of cloud contamination in the CO and $^{13}$CO J=2--1 lines are marked with arrows. Large gray points show radially-binned surfaces and small, light gray points represent individual measurements. The orange line in the CO emission surface show the exponentially-tapered power law fit (Table \ref{tab:emission_surf}).}
\label{fig:HD97048_fig}
\end{figure*}

Radial intensity profiles were generated using the \texttt{radial\_profile} function in the \texttt{GoFish} python package \citep{Teague19JOSS} to deproject the zeroth moment maps. For line emission originating from elevated disk layers like CO, we must consider its emitting surface during the deprojection process. Following \citet{LawMAPSIII}, we deprojected radial profiles using the derived surfaces listed in Table \ref{tab:emission_surf} for all disks. Radial profiles were generated using azimuthal averages, except for those disks showing substantial cloud obscuration, where we used asymmetric wedges to avoid regions of cloud contamination. This was necessary for WaOph~6 and Sz~91, where we used $\pm$55$^{\circ}$ and $\pm$90$^{\circ}$ wedges in the southern and northern parts of the disks, respectively. We also used a $\pm$30$^{\circ}$ wedge in DoAr~25 along the disk major axis, due to its highly inclined nature, to avoid including the shadowed disk midplane. Figure \ref{fig:line_emission_radial_profiles} shows the resultant radial profiles. For further discussion of the zeroth moment map and radial intensity profile generation process, see Sections 2.2 and 2.3, respectively, in \citet[][]{LawMAPSIII}.

To measure the radial extent of CO line emission, we calculated the disk size (R$_{\rm{CO}}$) as the radius which encloses 90\% of the total flux \citep[e.g.,][]{Tripathi17, Ansdell18}. This definition also allows for a direct comparison with the size of the MAPS disks derived in the same way in \citet{LawMAPSIII}. However, R$_{\rm{CO}}$ does not always reflect the outermost portion of CO emission in a disk, especially for those sources with low-intensity, plateau-like emission at large radii, e.g., CI~Tau, DM~Tau. Instead, to measure the outermost edge (R$_{\rm{edge}}$) of the CO gas disk, we computed the radius which encloses 99\% of the total disk flux. Both measurements were performed using the radial profiles in Figure \ref{fig:line_emission_radial_profiles}. Table \ref{tab:disksize} shows the CO gas disk size measurements and both R$_{\rm{CO}}$ (gray) and R$_{\rm{edge}}$ (red) are marked in Figure \ref{fig:line_emission_radial_profiles}. Overall, we find R$_{\rm{CO}}$ values that are generally consistent with those reported in \citet{Long22}. We do, however, find considerably smaller R$_{\rm{CO}}$ values for the V4046~Sgr (${\gtrsim}$20\%) and DoAr~25 (${\gtrsim}$30\%) disks. For the V4046~Sgr disk, this is likely driven by the coarse angular resolution (${\gtrsim}1^{\prime \prime}$) of the CO observations used by \citet{Long22}, while for the DoAr~25 disk, the ability to draw a wedge precisely along the CO emission surface to avoid confusion from the disk midplane likely leads to an improved estimate of R$_{\rm{CO}}$.


\section{Isovelocity Contours} \label{sec:appendix_isovel}

Figure \ref{fig:isovel_contours} shows the predicted isovelocity contours for CO line emission in representative channels in our sample. We show contours for only those radii where we were able to directly constrain the CO line emitting heights.

\section{Imaging and re-analysis of HD~97048 ALMA data} \label{sec:appendix_HD97048}

The CO J=2--1 emission surface of HD~97048 (CU~Cha) was extracted by \citet{Rich21} using archival ALMA data (PI: G. van der Plas, 2015.1.00192.S)\footnote{We note that the ALMA project code 2016.1.00826.S is incorrectly cited in \citet{Rich21}, but the authors instead used the CO J=2--1 transition from 2015.1.00192.S.}. However, the archival data does not provide continuum+line image cubes necessary for extracting temperatures (Section \ref{sec:calc_gas_temperatures}).

We re-imaged both the line-only and line+continuum CO data for this disk. Since the line-only data was taken from the pipeline-produced images, we also reprocessed this data to improve image quality. Since this ALMA program contained $^{13}$CO and C$^{18}$O J=2--1 isotopologue data, we also processed and imaged these line data. In CASA \texttt{v4.7.2} \citep{McMullin_etal_2007}, the 1.3~mm continuum was self-calibrated using two rounds of phase self-calibration, which was then applied to the continuum-subtracted line data. Both continuum and line imaging was performed with \texttt{tclean} with uniform weighting, which resulted in the 1.3~mm continuum image having a beam size of 0\farcs43 $\times$ 0\farcs21, PA=23.8$^{\circ}$ and an rms of 0.08~mJy/beam. The CO J=2--1 data had a beam size of 0\farcs45 $\times$ 0\farcs20 with PA=30$^{\circ}$, while the $^{13}$CO J=2--1 and C$^{18}$O J=2--1 data had beam sizes of 0\farcs42 $\times$ 0\farcs18 with PA=23$^{\circ}$. Typical line rms values were ${\approx}$5-9~mJy/beam. Figure \ref{fig:HD97048_fig} shows the 1.3~mm continuum image and the zeroth moment maps for CO, $^{13}$CO, and C$^{18}$O J=2--1 produced with \texttt{bettermoments} as in Appendix \ref{sec:appendix_disksize_profiles}.

As in Section \ref{sec:methods}, we used \texttt{disksurf} to extract emission surfaces for the CO J=2--1 and $^{13}$CO J=2--1 lines but were unable to derive line emitting heights for C$^{18}$O J=2--1. We find a CO emission surface that is consistent with the one derived in \citet{Rich21}. Due to the coarse and elongated beam size, it is possible that the CO and $^{13}$CO J=2--1 emission surfaces are modestly underestimated. However, we note that \citet{Pinte19Nat} found a $^{13}$CO J=3--2 emission height of 17~au at a radius of 130~au using a ${\approx}$0\farcs1 beam and similar surface extraction method. This closely agrees with the $^{13}$CO J=2--1 height that we derived at the same radius.

Radial and 2D temperature profiles were calculated using the line+continuum cubes, as in Section \ref{sec:gas_temperatures}, and are shown in Figures \ref{fig:figure_temp} and \ref{fig:2D_temp_surfaces}.

\begin{deluxetable}{lccc}[!t]
\tablecaption{Characteristic $z/r$ of CO Emission Surfaces \label{tab:char_zr_table}\vspace{-4mm}} 
\tablewidth{0pt}
\tablehead{
\colhead{Source} & \colhead{Line}  & \colhead{$r_{\rm{cutoff}}$ [au]} & \colhead{$z/r$} }
\startdata
\textbf{This work:}\\
HD~142666\tablenotemark{a} & J=2$-$1 & 51 & 0.20$^{+0.02}_{-0.05}$\\
MY~Lup & J=2$-$1 & 101 & 0.16$^{+0.01}_{-0.06}$\\
V4046~Sgr & J=3$-$2 & 114 & 0.24$^{+0.12}_{-0.07}$\\
HD~100546 & J=2$-$1 & 88 & 0.24$^{+0.07}_{-0.05}$\\
GW~Lup & J=2$-$1 & 150 & 0.21$^{+0.13}_{-0.02}$\\
WaOph~6 & J=2$-$1 & 110 & 0.19$^{+0.03}_{-0.05}$\\
DoAr~25 & J=2$-$1 & 177 & 0.25$^{+0.04}_{-0.06}$\\
Sz~91\tablenotemark{b} & J=3$-$2 & 108 & 0.22$^{+0.05}_{-0.10}$\\
CI~Tau & J=3$-$2 & 224 & 0.24$^{+0.07}_{-0.04}$\\
DM~Tau & J=2$-$1 & 204 & 0.53$^{+0.04}_{-0.09}$\\
\\\textbf{Literature:} & \\
IM~Lup\tablenotemark{a} & J=2$-$1 & 300 & 0.34$^{+0.08}_{-0.11}$\\
GM~Aur & J=2$-$1 & 479 & 0.35$^{+0.06}_{-0.11}$\\
AS~209 & J=2$-$1 & 173 & 0.17$^{+0.04}_{-0.07}$\\
HD~163296 & J=2$-$1 & 191 & 0.24$^{+0.06}_{-0.09}$\\
MWC~480 & J=2$-$1 & 401 & 0.22$^{+0.05}_{-0.11}$\\
HD~97048 & J=2$-$1 & 403 & 0.26$^{+0.05}_{-0.03}$\\
\enddata
\tablenotetext{a}{Cutoff radius manually adjusted.}
\tablenotetext{b}{Emission surface data were averaged starting at ${>}$50~au.}
\tablecomments{Literature sample composed of the disks around IM~Lup, GM~Aur, AS~209, HD~163296, and MWC~480 \citep{Law21}; and HD~97048 \citep{Rich21} with directly mapped CO line emission surfaces. Characteristic $z/r$ values are computed as the 50th percentile interior to r$_{\rm{cutoff}}$ and the uncertainties show the 16th to 84th percentile range.}
\end{deluxetable}

\section{Definition of Characteristic z/r of CO Emission Surfaces} \label{sec:app:definition_of_zr}

\begin{figure*}[!]
\centering
\includegraphics[width=\linewidth]{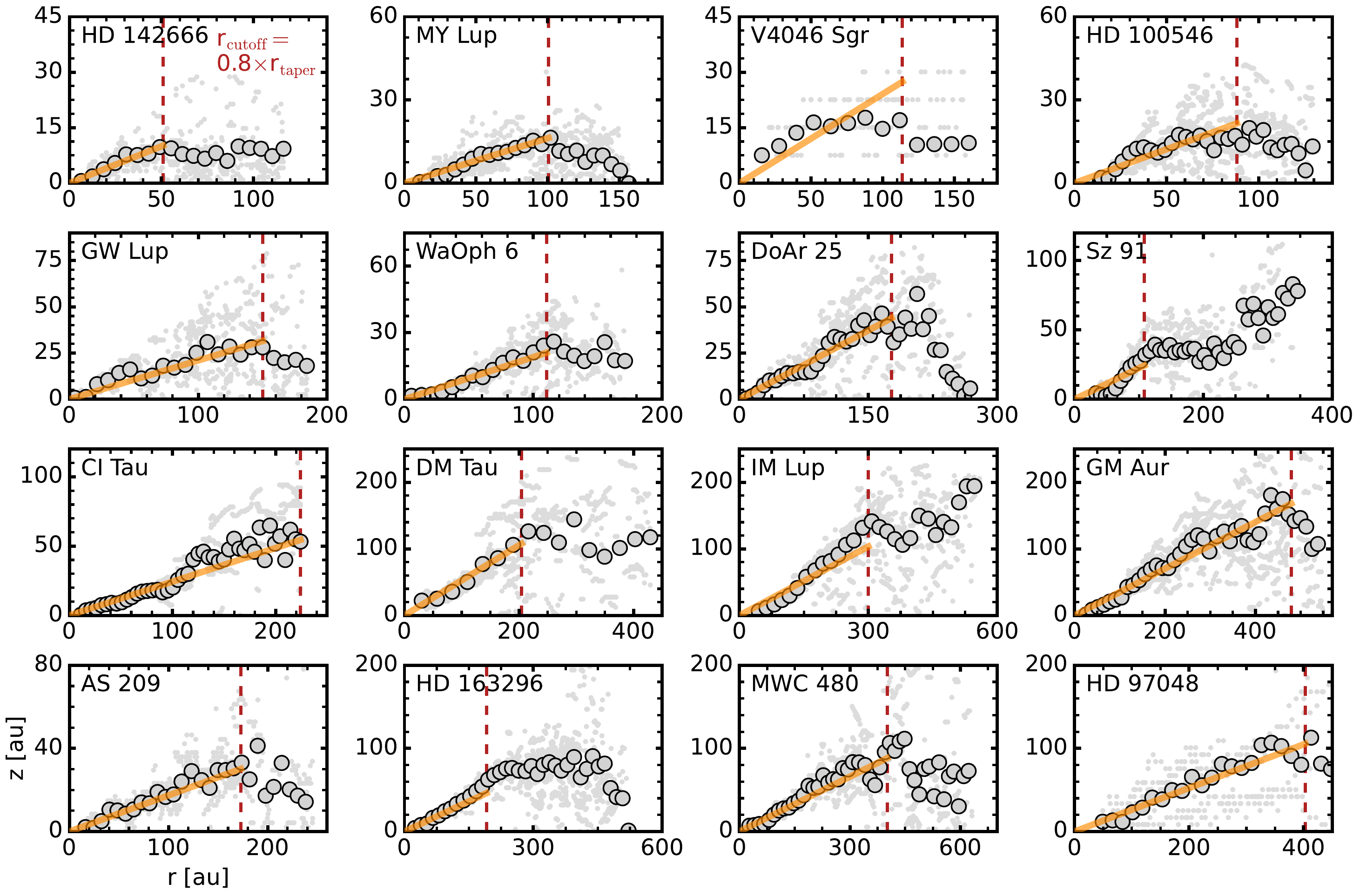}
\caption{Gallery for CO line emission surfaces for our sample and the IM~Lup, GM~Aur, AS~209, HD~163296, MWC~480 \citep{Law21} and HD~97048 disks \citep{Rich21}. Large gray points show radially-binned surfaces and small, light gray points represent individual measurements. Characteristic $z/r$ values are shown in orange and were computed as the mean $z/r$ for radii within r$_{\rm{cutoff}}=0.8\times$r$_{\rm{taper}}$, which is marked by a dashed red line.} 
\label{fig:zr_definition}
\end{figure*}

To enable a homogeneous comparison among sources, we required a characteristic $z/r$ for each CO line emission surface. We chose this $z/r$ to describe the inner rising portions of the line emission surfaces before the surfaces being to plateau and turnover due to, e.g., decreasing gas surface densities or insufficient observational line sensitivities. We defined this quantity as the mean $z/r$ computed from all points in the binned surfaces interior to a fixed cutoff radius. We chose r$_{\rm{cutoff}}=$0.8$\times$r$_{\rm{taper}}$, where r$_{\rm{taper}}$ is the fitted parameter from the exponentially-tapered power laws from Table \ref{tab:emission_surf}. We visually confirmed that 80\% of r$_{\rm{taper}}$ only included the rising part of the emission surfaces for all sources in our sample and in literature sources with directly mapped line emitting heights \citep{Law21, Rich21} with the exception of HD~142666 and IM~Lup. For these two disks, r$_{\rm{cutoff}}$ was manually chosen due to the lack of a clear turnover phase in either of their emission surfaces. Due to the relatively flat inner portion of the emission surface of the Sz~91 disk, we only averaged those points beyond 50~au when computing its characteristic $z/r$. We also calculated the 16th to 84th percentile range within r$_{\rm{cutoff}}$ as a proxy of the lower and upper flaring ranges, respectively, for each surface. Table \ref{tab:char_zr_table} lists the characteristic $z/r$, flaring ranges, and r$_{\rm{cutoff}}$ values for all sources in our sample and from the literature.

The characteristic $z/r$ is, by definition, constant and generally matches the binned surfaces well. However, at large radii, near r$_{\rm{cutoff}}$, this $z/r$ sometimes modestly underestimates the measured CO emission surface. This is the result of the flared nature of line emission surfaces and can be clearly in several sources, e.g., CI~Tau, Sz~91, IM~Lup, HD~163296, in Figure \ref{fig:zr_definition}.

\newpage
\clearpage


\bibliography{CO_surfaces}{}
\bibliographystyle{aasjournal}



\end{document}